\documentclass[a4paper,11pt]{article}
\usepackage{a4wide}
\usepackage{float}
\usepackage{graphicx}
\usepackage{amsmath}
\usepackage{xcolor}
\def\sech{\mathop{\rm sech}\nolimits}

\title{Whitham Shocks and Resonant Dispersive Shock Waves Governed by the Higher Order Korteweg-de Vries Equation}
\author{Saleh Baqer,\\
Department of Mathematics, Faculty of Science, \\
Kuwait University, Kuwait City 13060, Kuwait \and
Noel F. Smyth, \\
School of Mathematics, University of Edinburgh,\\
Edinburgh, Scotland, EH9 3FD, U.K. \\ and \\
School of Mathematics and Applied Statistics,\\
University of Wollongong,\\
Northfields Avenue, Wollongong, New South Wales, Australia, 2522.}
\date{}

\begin{document}

\maketitle

% Abstract

\begin{abstract}
 The addition of higher order asymptotic corrections to the Korteweg-de Vries equation results in the extended Korteweg-de Vries equation.  These higher order terms destabilise the dispersive shock wave solution, also termed an undular bore in fluid dynamics, and result in the emission of resonant radiation.  In broad terms, there are three possible dispersive shock wave regimes: radiating dispersive shock wave (RDSW), cross-over dispersive shock wave (CDSW) and travelling dispersive shock wave (TDSW).  While there are existing solutions for the RDSW and TDSW regimes obtained using modulation theory, there is no existing solution for the CDSW regime.  Modulation theory and the associated concept of a Whitham shock are used to obtain this CDSW solution.  In addition, it is found that the resonant wavetrain emitted by the extended Korteweg-de Vries equation with water wave coefficients has a minimal amplitude.  This minimal amplitude is explained based on the developed Whitham modulation theory.
\end{abstract}

{\textit{This paper is dedicated to Gerald B. Whitham, FRS whose prophetic speculations on dispersive shocks in the 1960s and 1970s have been
verified by this and other publications.}}

\section{Introduction}
\label{s:intro}

A solitary wave is the standard solution of nonlinear, dispersive wave equations, intensively studied since the 
1960s, when it was found that certain solitary wave supporting equations are integrable via the method of inverse scattering \cite{whitham}.  A solitary wave is termed a soliton as for such equations solitary waves interact cleanly, with no change of form (other than a phase shift), the word soliton then chosen due to its connotation with interacting sub-atomic particles.  The dispersive shock wave (DSW), also termed an undular bore in fluid mechanics, is another generic  solution of nonlinear, dispersive wave equations which has been receiving increased study.  In its standard form a DSW is a non-steady modulated wavetrain, consisting of solitary waves at one edge and linear dispersive waves at the opposite edge, connecting two distinct flow states. The separation between these two edges continuously increases. 
%\end{fmtext}
DSWs are formed due to the dispersive resolution of a discontinuity, with the simplest initial condition generating one being a step.  Undular bores were first observed as the tidal bores which occur in coastal areas of strong tide and appropriate topography, for example in Australia, Brazil, Canada, China, France, the United Kingdom and the United States.  However, DSWs are more widely observed in nature, with applications to water waves \cite{baines,esler}, meteorology \cite{christie,clarke,anne}, oceanography \cite{nwshelf},
geophysics \cite{scott1,scott2,hoefer2}, solid mechanics \cite{hooper}, nonlinear optics \cite{fleischer2,fleischer,elopt,colloid,nemboreel,nemborephysd,saleh,nonlocaltolocal}, Bose-Einstein condensates
\cite{bose} and Fermionic fluids \cite{hoefer1}.

DSW solutions are found using Whitham modulation theory \cite{whitham,mod1,modproc}.  Whitham modulation theory is an asymptotic technique used to study slowly varying periodic wave solutions of dispersive wave equations and gives partial differential equations, termed modulation equations, for the slowly varying parameters of the wavetrain, for instance its amplitude, wavelength and mean height.  In the case for which these modulation equations form a hyperbolic system the underlying periodic wavetrain is modulationally stable.  A simple wave solution of these hyperbolic modulation equations is then the DSW solution of the underlying equation \cite{gur,bengt}.  Such simple wave solutions are easy to determine if the modulation equations can be set in Riemann invariant form, which is guaranteed if the underlying equation is integrable \cite{flash}, but is difficult for non-integrable equations.  It has been found that in the solitary wave and linear wave limits modulation equations have degenerate forms, from which it was shown that the solitary wave and linear wave edges of a DSW can be determined without knowledge of the full modulation equations or their Riemann invariant form \cite{el2}.  An extensive review of DSWs, their physical applications and the connections between Whitham modulation equations and DSWs can be found in \cite{elreview}.

\begin{figure}[!ht]
    \centering
    \includegraphics[angle=270,width=0.45\textwidth]{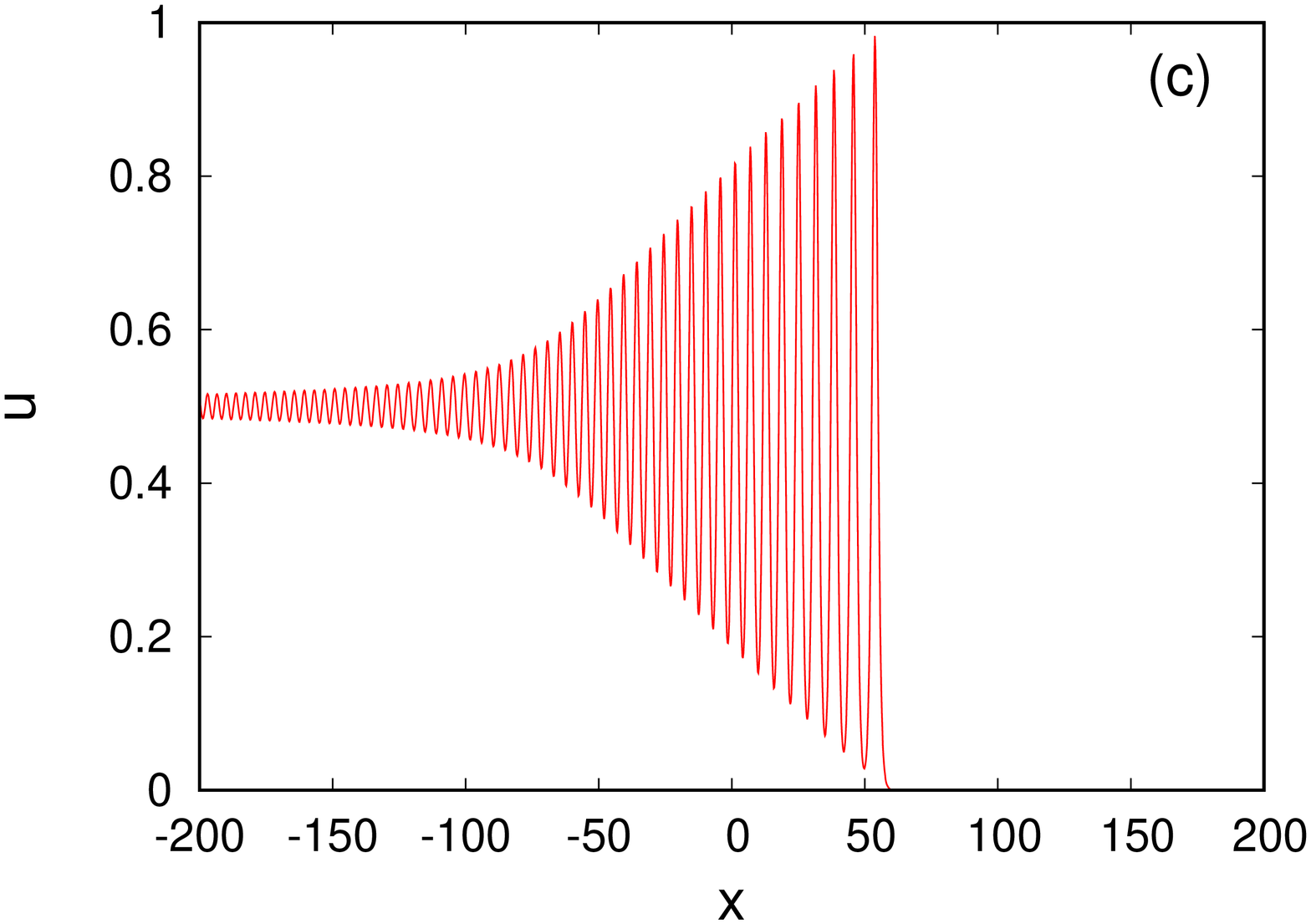}
    \includegraphics[angle=270,width=0.45\textwidth]{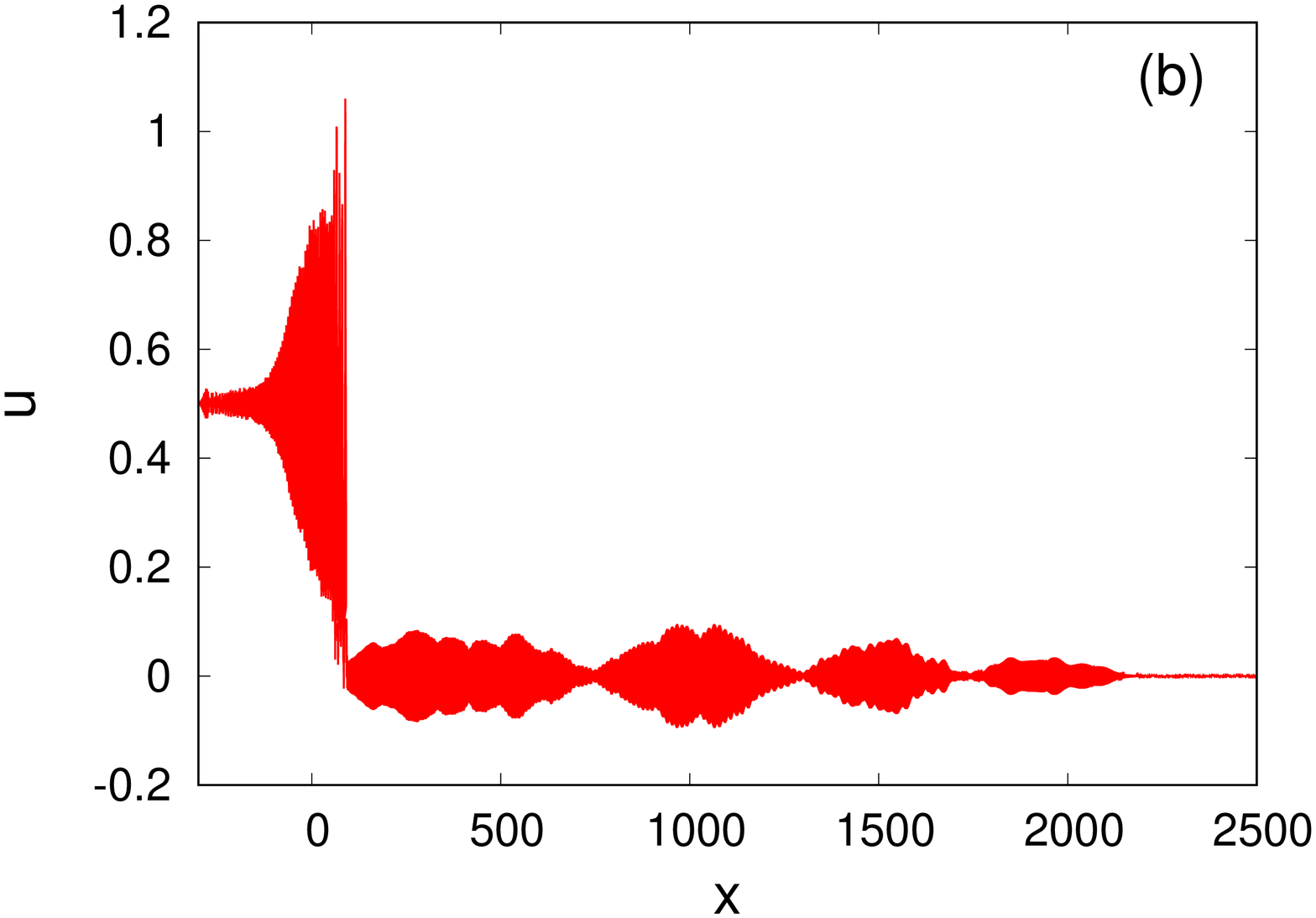}
    \includegraphics[angle=270,width=0.45\textwidth]{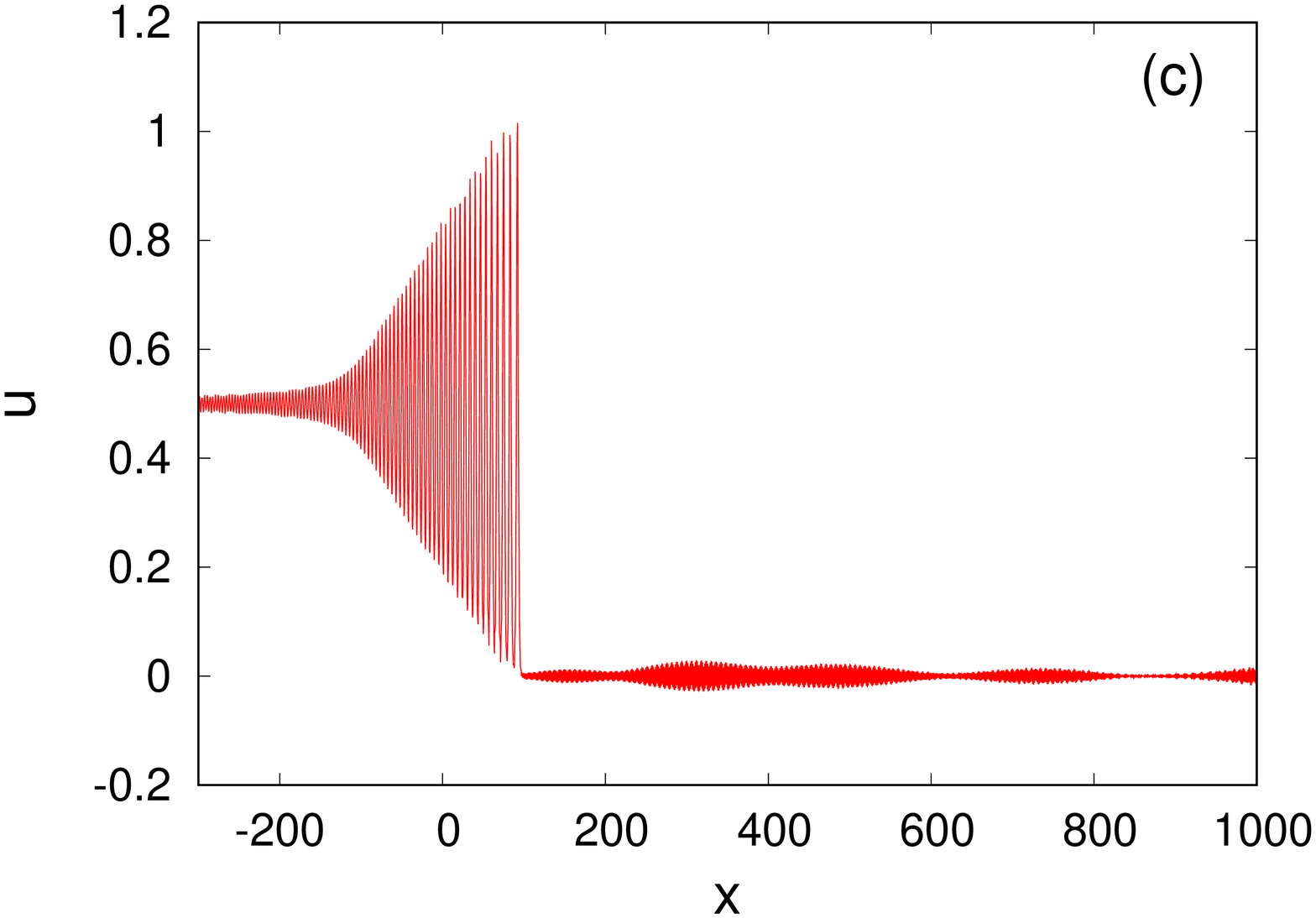}\includegraphics[angle=270,width=0.45\textwidth]{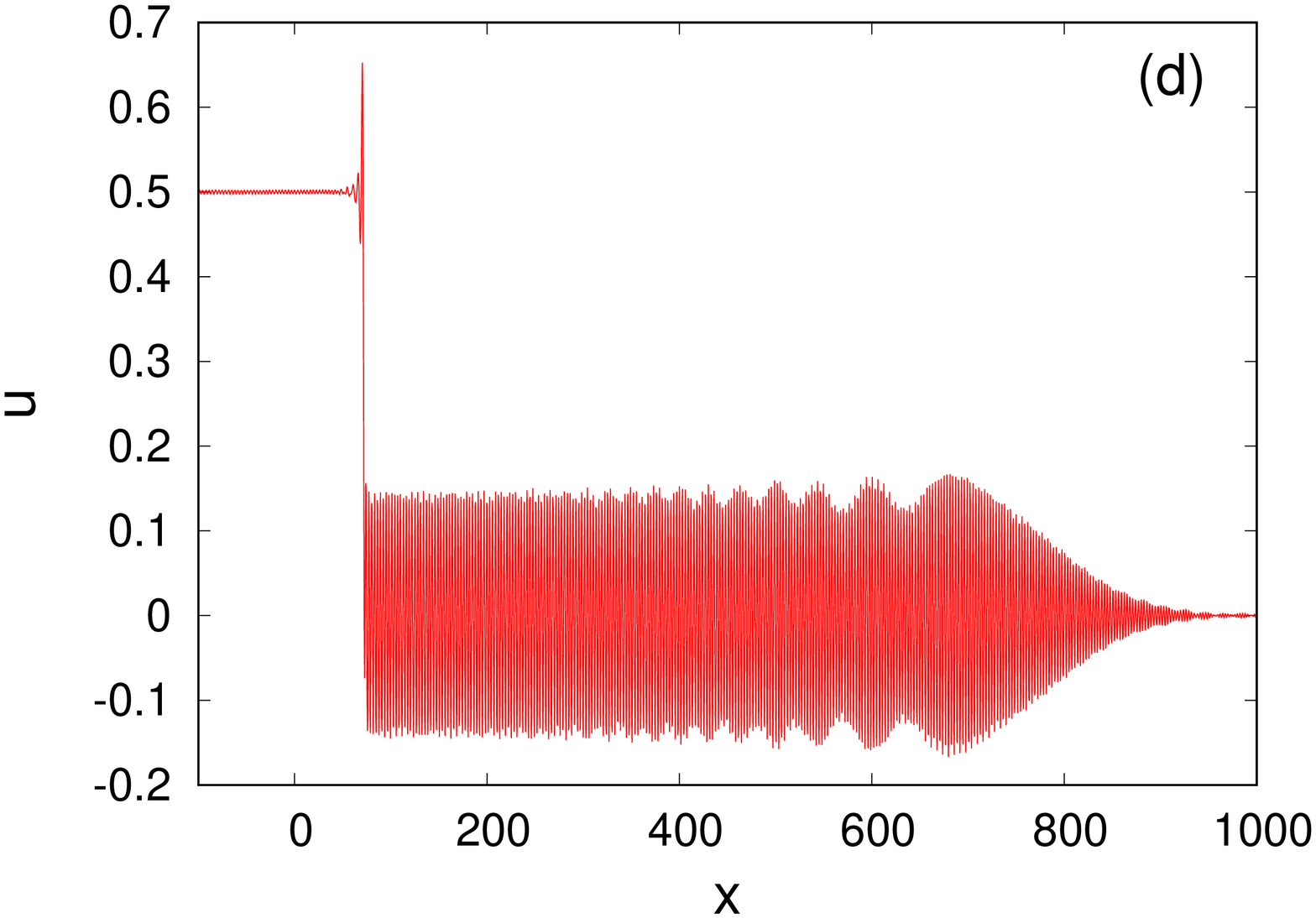}    \caption{DSW types. (a) classical KdV DSW with $c_{i}=0$, $i=1,\ldots, 4$, (b) non-classical CDSW with $c_{1}=-1$, $c_{2}=c_{3}=1$, $c_{4}=0.4$, (c) non-classical RDSW with $c_{1}=-1$, $c_{2}=c_{3}=1$, $c_{4}=0.3$, (d) non-classical TDSW with $c_{1}=-1$, $c_{2}=c_{3}=1$, $c_{4}=2.0$ . In these figures, $t=50$, $\epsilon = 0.15$ with $u_{-}=0.5$ and $u_{+}=0$.}
     \label{f:cdsw}
\end{figure}

This standard view of DSWs substantially alters for nonlinear, dispersive wave equations with non-convex dispersion.  This non-convex dispersion allows dispersive radiation to be in resonance with solitary waves, so that
the solitary wave sheds dispersive radiation and so is nonlocal  
\cite{kawahara,boyd,grim93,radiating_sol1,radiating_sol2,kivshar}.
The solitary wave then radiates away.  As a DSW is a modulated wavetrain with solitary waves at one edge, non-convex dispersion means that a DSW can also be resonant, shedding radiation, with the individual waves of the DSW being in resonance, not just the solitary wave at one edge.  As the DSW connects distinct flow states, it does not radiate away as the mass shed in radiation is replaced by that of the flow state into which it expands.  The addition of higher order dispersion to
standard nonlinear, dispersive wave equations, such as 
the Korteweg-de Vries equation \cite{markkaw,resekdv} and the nonlinear Schr\"odinger (NLS) equation \cite{trillores,trillo1,trilloresfour,trilloresnature,trilloreslossbore}, results in their DSWs being resonant.  
In addition, the DSW solution of the equations governing nonlinear optical beam propagation in nematic liquid crystals can be resonant or non-resonant \cite{nemboreel,saleh,nonlocaltolocal, salehthesis}, depending on 
the size of the jump of beam power across it.  The extended Korteweg-de Vries (eKdV) equation is an extension of the standard Korteweg-de Vries (KdV) equation for which the next higher order dispersive, nonlinear and nonlinear-dispersive terms are included in the asymptotic expansion \cite {whitham} which yields the KdV equation from more general equations, such as the water wave equations \cite{ekdv}.  The eKdV DSW solution, undular bore, is resonant \cite{resekdv}, with the higher order, fifth order, dispersion being a major driver of this \cite{markkaw}.  This resonance has a profound effect on the structure of the DSW, with the classical structure outlined above destroyed if the higher order terms are strong enough. Three regimes have been identified, radiating dispersive shock wave (RDSW), see Figure \ref{f:cdsw}(c), crossover dispersive shock wave (CDSW), see Figure \ref{f:cdsw}(b), and travelling dispersive shock wave (TDSW), see Figure \ref{f:cdsw}(d).  An RDSW occurs when the effect of the higher order terms is weak and so is a perturbed, radiating version of the classical DSW.  Above a threshold the resonance completely destroys the classical DSW structure, leaving a non-oscillatory jump between the initial levels of the step generating it.  Essentially, the classical DSW is radiated away by the dominant resonant wavetrain.  Between these two regimes, there is the CDSW, illustrated in Figure \ref{f:cdsw}(b).  The classical DSW, illustrated in Figure \ref{f:cdsw}(a) by the KdV DSW, is destabilised by the shed resonant wavetrain, which results in a structure similar to that for the focusing NLS equation with the amplitude ordered classical DSW of Figure \ref{f:cdsw}(a) replaced by an amplitude disordered CDSW of nearly constant average amplitude at its leading edge, but with a rapid decrease to the level behind
over its trailing edge, as seen in Figure \ref{f:cdsw}(b), so that a CDSW has a nearly constant amplitude over much of its length \cite{tovbis}.  That is, both the unstable focusing NLS DSW and the CDSW are large genus wavetrains (multi-phase wavetrains) that result in the long term formation of a so-called soliton gas (a large number of randomly interacting solitary waves) \cite{tovbis, gennadygas, thibaultgas}.  However, the latter sheds resonant radiation due to the non-convex dispersion effect.   The connection between a soliton gas and a CDSW deserves more detailed study, which is beyond the present work .
%The genus number becomes very large as the time evolves.}
%This unstable CDSW regime is illustrated in Figures \ref{f:cdsw}(a) and (b).  
Figure \ref{f:cdsw}(d) displays the time evolution of the resonant wavetrain propagating ahead of the CDSW to emphasise details of its modulational instability. This instability is analogous to the Benjamin-Feir instability observed in water waves as sideband perturbations \cite{whitham,BF}, see \cite{vandyke} for experimental photographs of this phenomenon.  The underlying high frequency resonant wavetrain with its unstable amplitude modulation is clear. 

The eKdV equation RDSW can be found as a perturbation of the KdV equation DSW \cite{perturbkdv}.  The TDSW solution has also been found as a TDSW is essentially a jump between two levels connected to the resonant wavetrain \cite{markkaw,pat,patjump}.  The present work will determine the eKdV equation CDSW solution based on Whitham modulation theory and the associated concept of a Whitham shock, which is a shock in the modulation variables \cite{modproc,whitham,patjump} and links the CDSW to the resonant wavetrain.  When Whitham developed modulation theory he speculated on the role of shocks in the case for which the modulation equations were hyperbolic, so that the underlying periodic wavetrain is stable \cite{modproc}, but did not explore the topic in extensive detail. The use of shocks in hyperbolic modulation equations has only recently been developed \cite{patjump, sergeyshockfront}, with applications to the Kawahara equation \cite{patjump}, the fifth order KdV equation \cite{patjump} and the BBM equation \cite{sergeybbm}.  Finally, it has been previously found that there is a resonant wave amplitude minimum for the CDSW solution of the eKdV equation with water wave coefficients \cite{resekdv}.  This amplitude minimum is deduced as a result of the developed modulation theory.

\section{Extended Korteweg-de Vries equation}

The Korteweg-de Vries (KdV) equation arises as an asymptotic approximation to more general equations in the long wavelength, weakly nonlinear limit when weak nonlinearity is balanced with weak dispersion \cite{whitham}.  If this asymptotic expansion is taken to one order beyond the KdV approximation, the result is the extended Korteweg-de Vries (eKdV) equation \cite{ekdv}
\begin{equation}
 u_{t} + 6uu_{x} + u_{xxx} + \epsilon \left[ c_{1} u^{2}u_{x} + c_{2} u_{x}u_{xx} + c_{3} uu_{xxx} + c_{4} u_{xxxxx} \right] = 0.
 \label{e:ekdv}
\end{equation}
Here, $\epsilon$ is a parameter measuring the weak nonlinearity, which for water waves is the ratio of the wave amplitude to the undisturbed fluid depth.
In the particular case of surface water waves the higher order coefficients are $c_{1} = -3/2$, $c_{2} = 23/4$, $c_{3} = 5/2$ and $c_{4} = 19/40$ \cite{ekdv}.  The eKdV equation also arises in the nonlinear optics of coherent beam propagation in nematic liquid crystals \cite{nonlocaltolocal}, for which $\epsilon$ is the ratio of the optical beam amplitude above a background level to this background level, and nonlinear elasticity \cite{hooper}.  The higher order coefficients $c_{i}$, $i=1,\ldots,4$, are involved in this case and can be found in the original work.  The Kawahara equation is the special case for which there is only fifth order dispersion \cite{kawahara}, $c_{i}=0$, $i=1,2,3$, and $c_{4} \ne 0$,
\begin{equation}
 u_{t} + 6uu_{x} + u_{xxx} + \epsilon c_{4} u_{xxxxx} = 0.
 \label{e:kawahara}
\end{equation}
The Kawahara equation arises for gravity-capillary waves for which the Bond number is near $1/3$ \cite{markkaw}.
In the present work, to generate a DSW the step initial condition
\begin{equation}
 u(x,0) = \left\{ \begin{array}{cc}
                   u_{+}, & x > 0\\
                   u_{-}, & x<0
                  \end{array}
          \right. 
\label{e:ic}          
\end{equation}
will be used, with $u_{-} > u_{+}$.  

The eKdV equation has the mass conservation equation
\begin{equation}
 \frac{\partial}{\partial t} u + \frac{\partial}{\partial x} \left[ 3u^{2} + u_{xx} + \epsilon \left( \frac{1}{3} c_{1} u^{3} + c_{3}uu_{xx} + \frac{1}{2} \left( c_{2} - c_{3} \right) u_{x}^{2} + c_{4} u_{xxxx} \right) \right] = 0 .
 \label{e:ekdvmass}
\end{equation}
The eKdV energy conservation is less straightforward to derive \cite{nonlocaltolocal}.  On multiplying the eKdV equation (\ref{e:ekdv}) by $u$ and integrating by parts gives
\begin{eqnarray}
 & & \frac{\partial}{\partial t} \frac{1}{2}u^{2} + \frac{\partial}{\partial x} \left[ 2u^{3} + uu_{xx}
 - \frac{1}{2}u_{x}^{2} + \epsilon \left( \frac{1}{4} c_{1} u^{4} + \frac{1}{2} c_{2} u^{2}u_{xx} + c_{4} uu_{xxxx} - c_{4} u_{x}u_{xxx} \right. \right. \nonumber \\
 & & \left. \left. \mbox{} + \frac{1}{2} c_{4} u_{xx}^{2} \right) \right] + \epsilon \left( c_{3} - \frac{1}{2} c_{2} \right) u^{2}u_{xxx} = 0.
  \label{e:energycons}
\end{eqnarray}
To set the final term in conservation form we use the fact that $\epsilon$ is small.  The first order equation, the KdV equation ($\epsilon = 0$), gives
\begin{equation}
 \frac{\partial}{\partial t} u^{3} = -3u^{2} \left( 6uu_{x} + u_{xxx} \right) = -\frac{\partial}{\partial x} \frac{9}{2}u^{4} - 3u^{2}u_{xxx},
 \label{e:q3}
\end{equation}
so that (\ref{e:energycons}) becomes the eKdV energy conservation equation
\begin{eqnarray}
 & & \frac{\partial}{\partial t} \left[ \frac{1}{2}u^{2} - \frac{1}{3} \epsilon
  \left(c_{3} - \frac{1}{2} c_{2}\right) u^{3} \right] + \frac{\partial}{\partial x} \left[ 2u^{3} + uu_{xx}
 - \frac{1}{2} u_{x}^{2} + \epsilon \left( \frac{1}{4} c_{1} u^{4}
 \right. \right. \nonumber \\
 & & \left. \left. \mbox{} + \frac{1}{2} c_{2}u^{2}u_{xx} + c_{4}uu_{xxxx}
 - c_{4}u_{x}u_{xxx} + \frac{1}{2} c_{4} u_{xx}^{2} 
 - \frac{3}{2} \left( c_{3} - \frac{1}{2} c_{2} \right) u^{4} \right) \right] = 0,
 \label{e:energyconsf}
\end{eqnarray}
which is valid to $O(\epsilon^{2})$.  We note that if $c_{2} = 2c_{3}$, then this energy conservation law is exact.  However, this relation does not hold for the eKdV equation with the water wave coefficients.

The mass (\ref{e:ekdvmass}) and energy (\ref{e:energyconsf}) conservation equations are of the form
\begin{equation}
 \frac{\partial P_{m}}{\partial t} + \frac{\partial Q_{m}}{\partial x} = 0 , \quad \frac{\partial P_{e}}{\partial t} + \frac{\partial Q_{e}}{\partial x} = 0,
 \label{e:consform}
\end{equation}
where $P_{m}$, $P_{e}$ and $Q_{m}$, $Q_{e}$ are the mass and energy densities and fluxes, 
\begin{eqnarray}
  P_{m} & = & u, \label{e:pm} \\
  P_{e} & = & \frac{1}{2}u^{2} - \frac{1}{3} \epsilon
  \left(c_{3} - \frac{1}{2} c_{2}\right) u^{3},
  \label{e:pe} \\
  Q_{m} & = & 3u^{2} + u_{xx} + \epsilon \left( \frac{1}{3} c_{1} u^{3} + c_{3}uu_{xx} + \frac{1}{2} \left( c_{2} - c_{3} \right) u_{x}^{2} + c_{4} u_{xxxx}
  \right), \label{e:qm} \\
  Q_{e} & = & 2u^{3} + uu_{xx}
 - \frac{1}{2} u_{x}^{2} + \epsilon \left[ \frac{1}{4} c_{1} u^{4}
  + \frac{1}{2} c_{2}u^{2}u_{xx} + c_{4}uu_{xxxx}
 - c_{4}u_{x}u_{xxx} + \frac{1}{2} c_{4} u_{xx}^{2} \right.
 \nonumber \\
 & & \left. \mbox{} + \frac{3}{2} \left( \frac{1}{2} c_{2} - c_{3} \right) u^{4} \right], \label{e:qe}
\end{eqnarray}
respectively. 

\section{Resonant wave Stokes expansion}

It can be seen from Figure \ref{f:cdsw} that the amplitude of the resonant wavetrain generated by a CDSW is small, certainly smaller than the amplitude of the DSW itself.  The resonant wavetrain can then be taken as a Stokes wave expansion of the form 
\begin{equation}
 u_{r} = \bar{u}_{r} +a_{r} \cos{\theta_{r}} + a^{2}_{r} u_{2}\cos{2\theta_{r}}+O(a^3_{r}), \quad
\omega_{r} (k_{r},a_{r}) =\omega_{0}+a_{r} \omega_{1}+a^{2}_{r} \omega_{2}+O(a^3_{r}),
\label{e:stokes}
\end{equation}
where $\bar{u}_{r}$ is the mean height of the wavetrain.
Here, $a_{r}$ is the amplitude of the resonant wavetrain and $\omega_{r}$ and $k_{r}$ are its frequency and wavenumber, with $\theta_{r} = k_{r}x - \omega_{r} t$.
Substituting this Stokes wave expansion into the eKdV equation (\ref{e:ekdv}), separating out the resulting equation at $O(\epsilon^{n})$, $n=0,1,2$,
and eliminating secular terms gives
\begin{eqnarray}
& & \omega_{0} = (6\bar{u}_{r}+\epsilon c_{1}\bar{u}^{2}_{r})k_{r}-\left(1+\epsilon c_{3}\bar{u}_{r}\right)k^3_{r} + \epsilon c_{4} k^5_{r}, \quad \omega_{1}=0, \label{e:omega0} \\
& &  \omega_{2} = \frac{36+24\epsilon c_1\bar{u}_{r} - \epsilon \left(48c_3-6c_1-6c_2\right)k^{2}_{r}}{24k_{r} - \epsilon\left(120c_4k^3_{r}-24c_3\bar{u}_{r}k_{r}\right)}+O\left(\epsilon^2\right)
\nonumber \\
& & = \frac{3}{2k_{r}}+\epsilon\left[\frac{1}{4}\left( c_{1}+c_2 -8c_{3}+ 30 c_{4}\right)k_{r} + \left( c_{1} - \frac{3}{2}c_{3} \right) \frac{\bar{u}_{r}}{k_{r}}\right]+O\left(\epsilon^2\right) \label{e:omega2}
\end{eqnarray}
and
\begin{equation}
 u_{2} = \frac{6+2\epsilon c_{1}\bar{u}_{r}-\epsilon (c_{2}+c_3)k^2_{r}}{12k^2_{r} + 12\epsilon\left(c_{3}\bar{u}_{r} - 5c_4k^2_{r}\right)k^{2}_{r}}  =
 \frac{1}{2k^2_{r}} - \frac{1}{12}\epsilon\left[ c_2+c_3 - 30 c_{4} - 2\left(c_{1} - 3c_{3} \right) \frac{\bar{u}_{r}}{k^2_{r}}\right] + O\left(\epsilon^2\right). \label{e:u2}
\end{equation}
The reason that the Stokes wave coefficients are expanded for small $\epsilon$ in (\ref{e:omega0})--(\ref{e:u2}) is to make the Whitham modulation jump conditions that connect the bore with the resonant radiation ahead in the CDSW regime as simple as possible.

\section{CDSW equal amplitude approximation}

\begin{figure}[!ht]
    \centering
    \includegraphics[angle=270,width=0.45\textwidth]{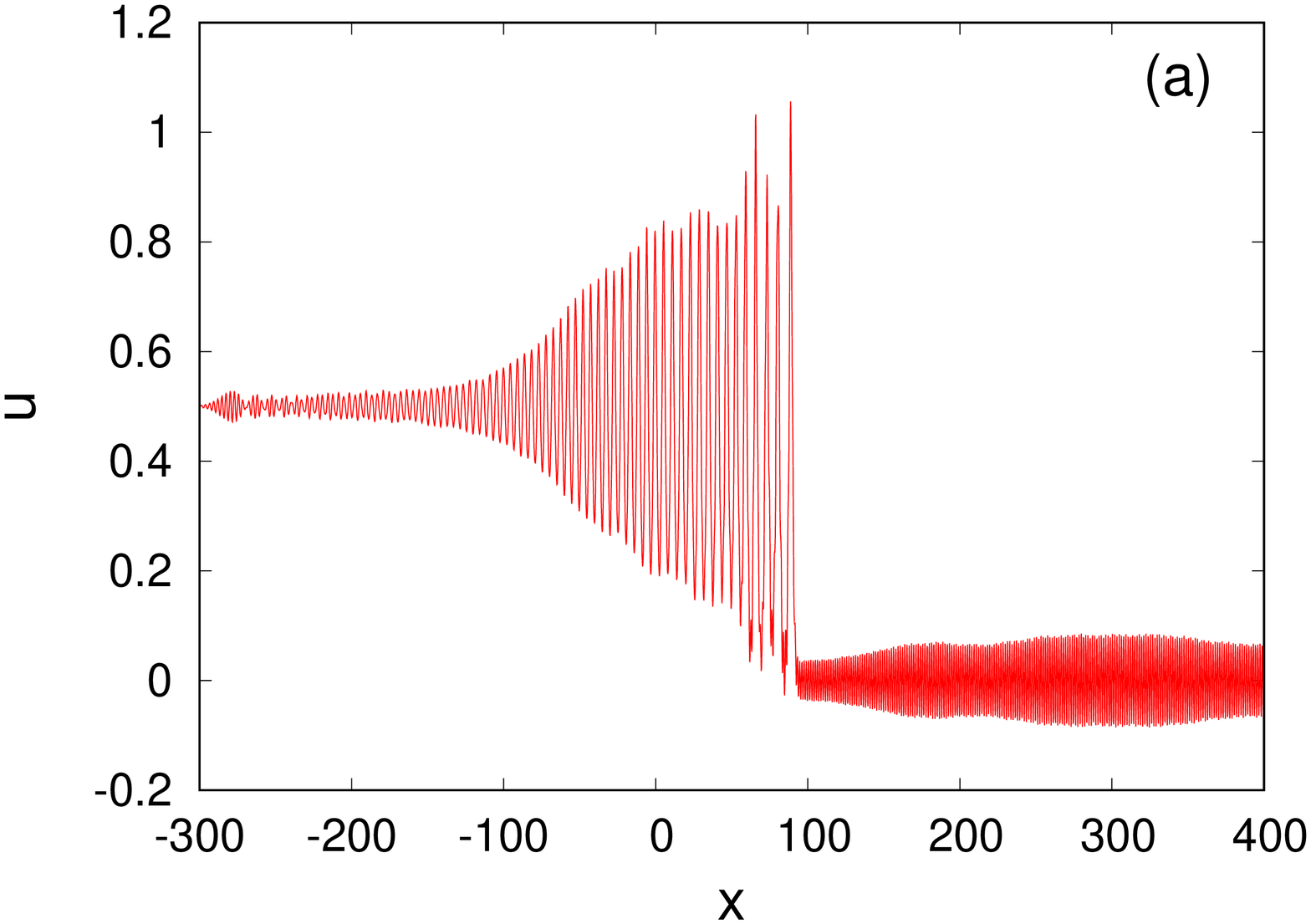}
    \includegraphics[width=0.68\textwidth]{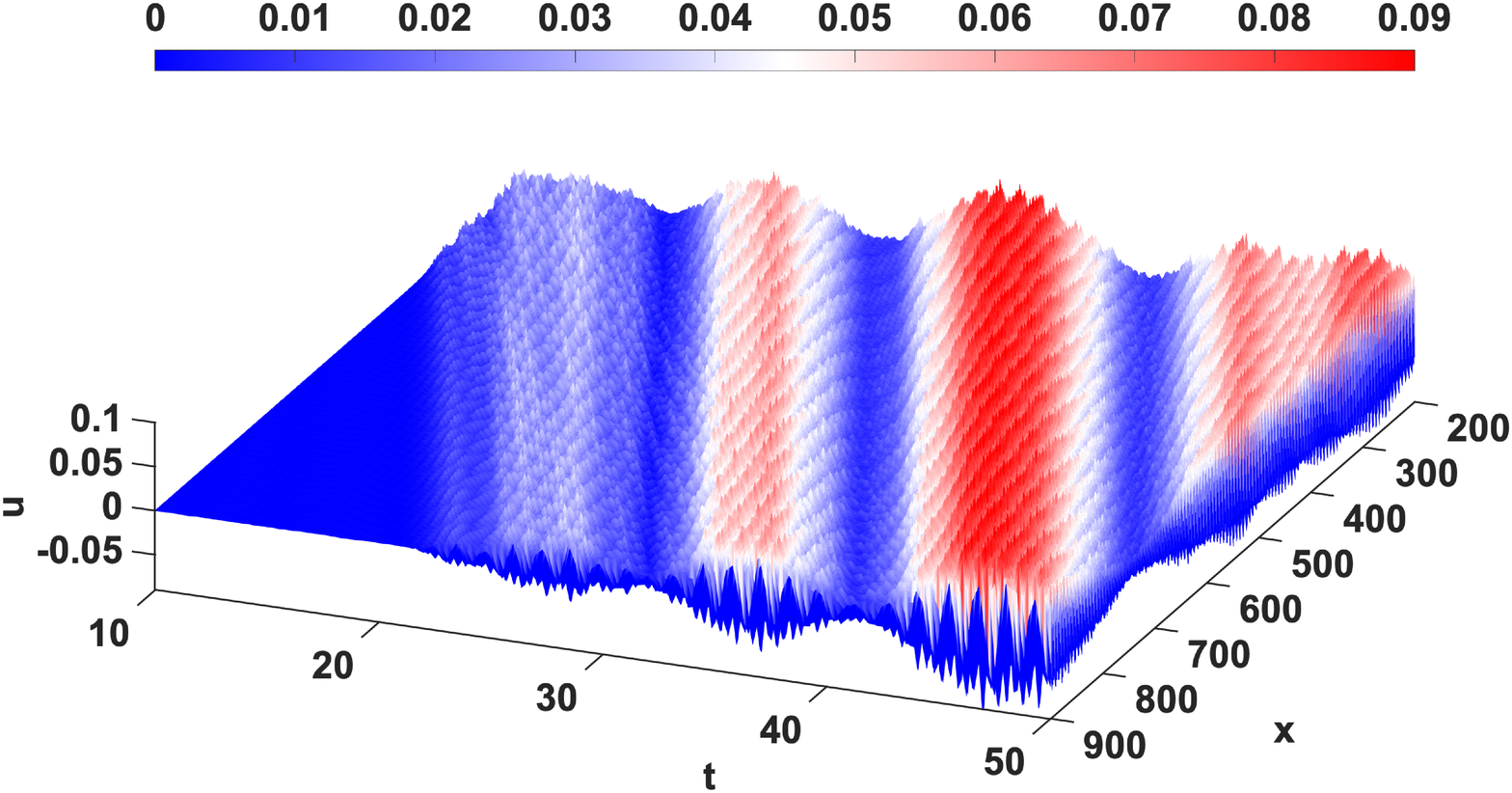}(b)
    \caption{Example CDSWs.  (a) Detail of CDSW at $t=50$, (b) evolution detail of resonant wavetrain.  In these figures, $c_{1}=-1$, $c_{2}=c_{3}=1$, $c_{4}=0.4$ with $\epsilon = 0.15$ and $u_{-}=0.5$, $u_{+}=0$.}
     \label{f:cdsw2}
\end{figure}

As seen from the example shown in Figure \ref{f:cdsw2} the eKdV CDSW is unstable, as is its generated resonant wavetrain, see Figure \ref{f:cdsw2}(b),  and does not exhibit the standard KdV DSW structure, as discussed in Section \ref{s:intro}. The amplitudes of the waves of the CDSW do not decrease monotonically from the leading to the trailing edges.  The amplitudes of the waves are random due to the instability, but distributed around a constant mean, with a rapid decrease to $u_{-}$ at the trailing edge \cite{tovbis}.
This broad structure of a CDSW can be exploited to obtain an approximate solution for it \cite{saleh,nonlocaltolocal,boreapprox}, which can then be linked to loss radiated in the resonant wavetrain \cite{saleh,nonlocaltolocal}.  In essence, the CDSW is approximated by a train of uniform solitary waves, with mass and energy conservation used to determine the amplitude and spacing of these waves.

The basis of the equal amplitude approximation is the solitary wave solution of the eKdV equation.  While there is no exact solitary wave solution of this equation, there is a perturbation solution based on $\epsilon$ small
\cite{perturbkdv}
\begin{equation}\label{eKdV_soliton}
    u_{s} = { \bar{u}_{s} + \left(a_{s}+\epsilon c_6a^2_{s}\right)\sech^{2}{w_{s}\theta_{s}} + \epsilon c_7 a^{2}_{s}\sech^{4}{w_{s}\theta_{s}} + O (\varepsilon^2) },
\end{equation}
with phase $\theta_{s} = x - U_{s}t$, inverse width $w_{s} = \sqrt{a_{s}/2}$ and velocity $U_{s} = 2a_{s}+ 4\epsilon c_{4} a^2_{s}+O(\epsilon^2)$, for which the new constants $c_{6}$ and $c_{7}$ are
%\begin{equation}\label{theta_s}
%    w_{s}=\sqrt{\frac{a_{s}}{2}},\quad \theta_{s}=x-U_{s}t,\quad U_s=2a_{s}+ 4\epsilon c_{4} a^2_{s}+O(\epsilon^2),
%\end{equation}
\begin{equation}\label{c6_c7}
     c_6=- \frac{1}{6}c_1 + \frac{1}{6}c_2 + \frac{2}{3}c_{3} - 5c_4,\quad  c_7 = \frac{1}{12}c_1 - \frac{1}{4}c_2 - \frac{1}{2}c_{3} + \frac{15}{2}c_{4}.
\end{equation}

Let us assume that in the CDSW regime the solution consists of the level ahead $u_{+}$, joining to a resonant wavetrain of amplitude $a_{r}$, wavenumber $k_{r}$ and mean height $\bar{u}_{r}$ (the frequency $\omega_{r}$ given by the Stokes wave dispersion relation (\ref{e:stokes})), followed by a uniform train of $N(t)$ equal solitary waves
of amplitude $a_{s}$ on a mean level $\bar{u}_{s}$, which then links to the level $u_{-}$ behind.  This matches the general form of the example CDSW solution shown in Figure \ref{f:cdsw}.

The eKdV mass and energy conservation equations (\ref{e:ekdvmass}) and (\ref{e:energyconsf}) are now integrated in $x$ from the level behind $u_{-}$ to the trailing edge of the resonant wavetrain, giving
\begin{eqnarray}
 N \int_{-\infty}^{\infty} u_{s} \: dx & = & \left[ 3u_{-}^{2} + \frac{1}{3} \epsilon c_{1} u_{-}^{3} - \bar{Q}_{mr}\right] t, \label{e:masscdsw} \\
          N \int_{-\infty}^{\infty} \left[ \frac{1}{2} u_{s}^{2} - \frac{1}{3} \epsilon \left( c_{2} - \frac{1}{2} c_{2} \right)u_{s}^{3} \right] \: dx & = & \left[ 2u_{-}^{3} + \frac{1}{4} \epsilon \left( c_{1} + 3c_{2} -                                         6 c_{3} \right) u_{-}^{4} - \bar{Q}_{er} \right] t. \label{e:energycdsw}
\end{eqnarray}
Here, $\bar{Q}_{mr}$ and $\bar{Q}_{er}$ are the mass and energy fluxes (\ref{e:qm}) and (\ref{e:qe}) 
%\begin{eqnarray}
%  Q_{m} & = & 3u^{2} + u_{xx} + \epsilon \left( \frac{1}{3} c_{1} u^{3} + c_{3}uu_{xx} + \frac{1}{2} \left( c_{2} - c_{3} \right) u_{x}^{2} + c_{4} u_{xxxx}
%  \right), \label{e:qm} \\
%  Q_{e} & = & 2u^{3} + uu_{xx}
% - \frac{1}{2} u_{x}^{2} + \epsilon \left[ \frac{1}{4} c_{1} u^{4}
%  + \frac{1}{2} c_{2}u^{2}u_{xx} + c_{4}uu_{xxxx}
% - c_{4}u_{x}u_{xxx} + \frac{1}{2} c_{4} u_{xx}^{2} \right.
% \nonumber \\
% & & \left. \mbox{} + \frac{3}{2} \left( \frac{1}{2} c_{2} - c_{3} \right) u^{4} %\right]. \label{e:qe}
%\end{eqnarray}
averaged over the Stokes wave (\ref{e:stokes}), respectively.  Dividing these mass and energy results and evaluating the mass and energy fluxes over the resonant wavetrain gives a relation linking the solitary wave amplitude $a_{s}$ and mean height $\bar{u}_{s}$ to the amplitude $a_{r}$, wavenumber $k_{r}$ and mean height $\bar{u}_{r}$ of the resonant wavetrain.  As this expression is involved, particularly the energy density, the relation is detailed in Appendix \ref{a:appendixa}.  Dividing the mass and energy results (\ref{e:masscdsw}) and (\ref{e:energycdsw}) gives the implicit relation
\begin{equation}
 \frac{\int_{-\infty}^{\infty} u_s \: dx}{ \int_{-\infty}^{\infty} \left[ \frac{1}{2} u_{s}^{2} - \frac{1}{3} \epsilon \left( c_{2} - \frac{1}{2} c_{2} \right)u_{s}^{3} \right] \: dx} = \frac{3u_{-}^{2} + \frac{1}{3} \epsilon c_{1} u_{-}^{3} - \bar{Q}_{mr}}{2u_{-}^{3} + \frac{1}{4} \epsilon \left( c_{1} + 3c_{2} -                                         6 c_{3} \right) u_{-}^{4} - \bar{Q}_{er}},
 \label{e:cdswampvel}
\end{equation}
which determines the amplitude of the CDSW.  

\section{Modulation theory for CDSW and resonant wavetrain}
\label{s:modulation}

The resonance condition between the CDSW and the resonant wavetrain is that the solitary wave velocity $U_{s}$ equals the Stokes wave velocity, so that
\begin{equation}
2a_{s}+ 4\epsilon c_{4} a^2_{s} = c = \frac{\omega_{0} + 
a^{2}_{r}\omega_{2}}{k_{r}},
\label{e:resonance}
\end{equation}
where $\omega_{0}$ and $\omega_{2}$ are given by the dispersion relation (\ref{e:stokes}).  

The example numerical solution of the eKdV equation in the CDSW regime shown in Figure \ref{f:cdsw} shows that it consists of five distinct regions.  The initial level
behind $u_{-}$ links to an unstable DSW, which then shows a sharp jump to the shed resonant wavetrain, which is here modelled by a Whitham shock.  The resonant wavetrain then transitions to the initial level ahead $u_{+}$ via a ``partial DSW'' \cite{pat,negative}.  A ``full DSW,'' as for the standard KdV DSW, is a transition between two uniform levels via a modulated wavetrain with solitary waves at one edge and linear waves at the opposite edge.  A partial DSW links two uniform wavetrains via a modulated wavetrain whose amplitude, wavenumber and
mean height are continuous at the two edges.  A full DSW is a limit of a partial DSW with no steady wavetrains at its edges.  The partial DSW smoothly raises the initial level ahead $u_{+}$ to the mean level of the resonant wavetrain $\bar{u}_{r}$, which then jumps discontinuously to the mean level of the CDSW $\bar{u}_{s}$ via the Whitham shock joining the resonant wavetrain to the CDSW.  To fully determine the partial DSW the Whitham modulation equations for the resonant wavetrain need to be calculated \cite{negative}. 
This results in an involved system of equations \cite{negative} when the jump conditions across the Whitham shock are included.  However, $\bar{u}_{r}$ is very close to $u_{+}$ \cite{pat} and to a good approximation can be set to $u_{+}$ and the extra accuracy obtained by the inclusion of the partial DSW is minimal.   Indeed, the mean level will be found to have minimal change over the Whitham shock.  

It can be seen from Figure \ref{f:cdsw} that the DSW and its associated resonant wavetrain are unstable in the CDSW regime.  The Whitham modulation equations can then be assumed to be elliptic \cite{whitham,modproc}, even though they have not been calculated.  Although this ellipticity conclusion comes from numerical observations and not from an exact calculation of the associated Whitham modulation system, it was analytically verified in \cite{pat} that the Whitham modulation system becomes fully elliptic when the stable Kawahara DSW regime approaches the CDSW regime.  Extending the modulation theory work of \cite{pat} to eKdV CDSWs is outside the scope of the present study and will not be dealt with here.  It is another topic for future study.

In theory elliptic systems do not have discontinuous solutions with shock waves.  However, Figure \ref{f:cdsw} shows that there is a clear rapid connection between the CDSW and the resonant wavetrain, which we shall model as a discontinuity.  We shall then model this connection by a Whitham shock based on the conservation of mass and energy across it as these quantities have to be conserved for any valid solution.
This approach was found to be successful in the study of the DSW solution of the nematic equations \cite{saleh, nonlocaltolocal}.  

Averaging the mass and energy conservation equations (\ref{e:consform}) over the resonant wavetrain ahead of the Whitham shock and the CDSW behind the shock gives 
the jump conditions across the shock.  As the CDSW is led by solitary waves, the Whitham shock velocity is equal to the solitary wave velocity $U_{s}$ \cite{whitham}.  The jump conditions are then 
\begin{equation}
 -U_{s} \left[\bar{P}_{mcdsw} - \bar{P}_{mr}\right] + \left[\bar{Q}_{mcdsw} - \bar{Q}_{mr}\right] = 0, \quad
 -U_{s} \left[\bar{P}_{ecdsw} - \bar{P}_{er}\right] + \left[\bar{Q}_{ecdsw} - \bar{Q}_{er}\right] = 0,
 \label{e:jump}
\end{equation}
where $\bar{P}_{mcdsw}$, $\bar{P}_{ecdsw}$ and $\bar{P}_{mr}$, $\bar{P}_{er}$ denote the mass and energy densities (\ref{e:pm}) and (\ref{e:pe})
%\begin{eqnarray}
%  P_{m} & = & u, \label{e:pm} \\
%  P_{e} & = & \frac{1}{2}u^{2} - \frac{1}{3} \epsilon
%  \left(c_{3} - \frac{1}{2} c_{2}\right) u^{3}
%  \label{e:pe}
%\end{eqnarray}
averaged over the CDSW regime and the Stokes wave, respectively.  Similarly, $\bar{Q}_{mcdsw}$, $\bar{Q}_{ecdsw}$ and $\bar{Q}_{ecdsw}$, $\bar{Q}_{er}$ denote the mass and energy fluxes (\ref{e:qm}) and (\ref{e:qe}) averaged over the bore in the CDSW regime and the Stokes wave, respectively.  Here, $U_{s}$ denotes the Whitham shock velocity.

The CDSW amplitude relation (\ref{e:cdswampvel}), the resonance condition (\ref{e:resonance}) and the jump conditions (\ref{e:jump}) give four relations for the 
five unknowns $a_{s}$, $a_{r}$, $\bar{u}_{s}$, $\bar{u}_{r}$ and $k_{r}$.  With the approximation that $\bar{u}_{r} = u_{+}$, this gives a complete system of equations.  This system of equations to determine the resonant CDSW were solved numerically using Newton's method.  It was found to be sufficient to keep terms up to the orders $O(\epsilon)$ and $O(a^{2}_{r})$ in the CDSW amplitude relation (\ref{e:cdswampvel}), the resonance condition (\ref{e:resonance}) and the mass jump condition of (\ref{e:jump}) in order to keep these equations as simple as possible.  We note that
the eKdV equation (\ref{e:ekdv}) as a reduction of the water wave equations is asymptotically valid to $O(\epsilon)$.  However, it was found to be essential to keep all terms in the Whitham energy jump condition of (\ref{e:jump}) in order to obtain good agreement with numerical solutions.  Indeed, Newton's method had a tendency to not converge if all energy terms were not included.  Including higher order terms in the CDSW amplitude relation (\ref{e:cdswampvel}), the resonance condition (\ref{e:resonance}) and the mass jump condition of (\ref{e:jump}) resulted in no graphical difference in the resulting solution.
Indeed, if the theoretical resonant wave amplitude is truncated to the linear value $A_{r} = a_{r}$, there is no observable difference in the comparisons with numerical solutions.
%of Figure \ref{f:kawahara}(a).
As the full detailed jump conditions,
particularly the energy jump condition, are then extensive, they are given in Appendix \ref{a:jump}.  A similar situation was encountered in the study of nematic CDSWs when the nonlocality of the nematic medium is decreased to the local optical limit \cite{nonlocaltolocal}.

\section{Comparisons with numerical solutions}
\label{s:numerical}

\begin{figure}[!ht]
    \centering
    \includegraphics[angle=270,width=0.47\textwidth]{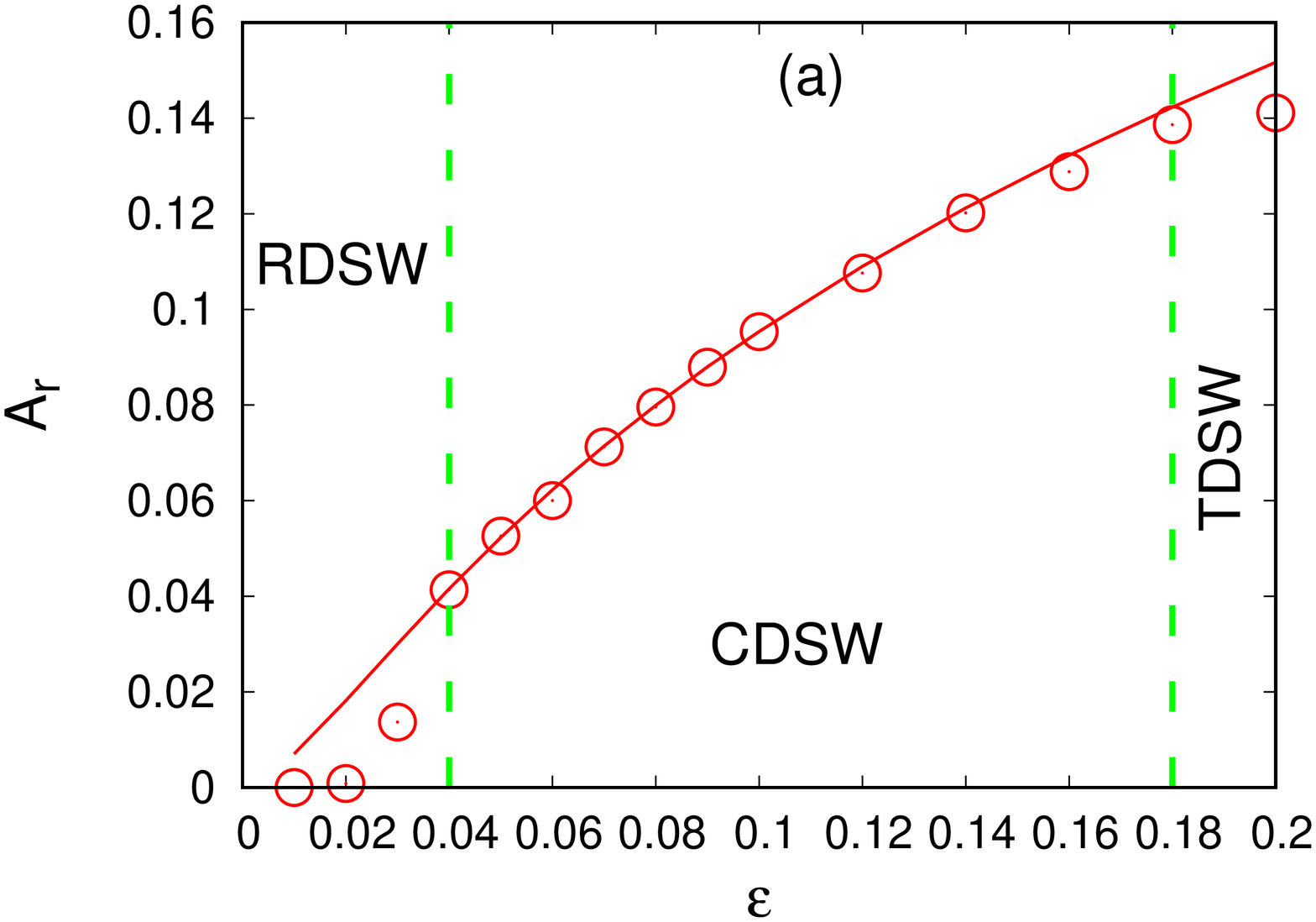}
    \includegraphics[angle=270,width=0.47\textwidth]{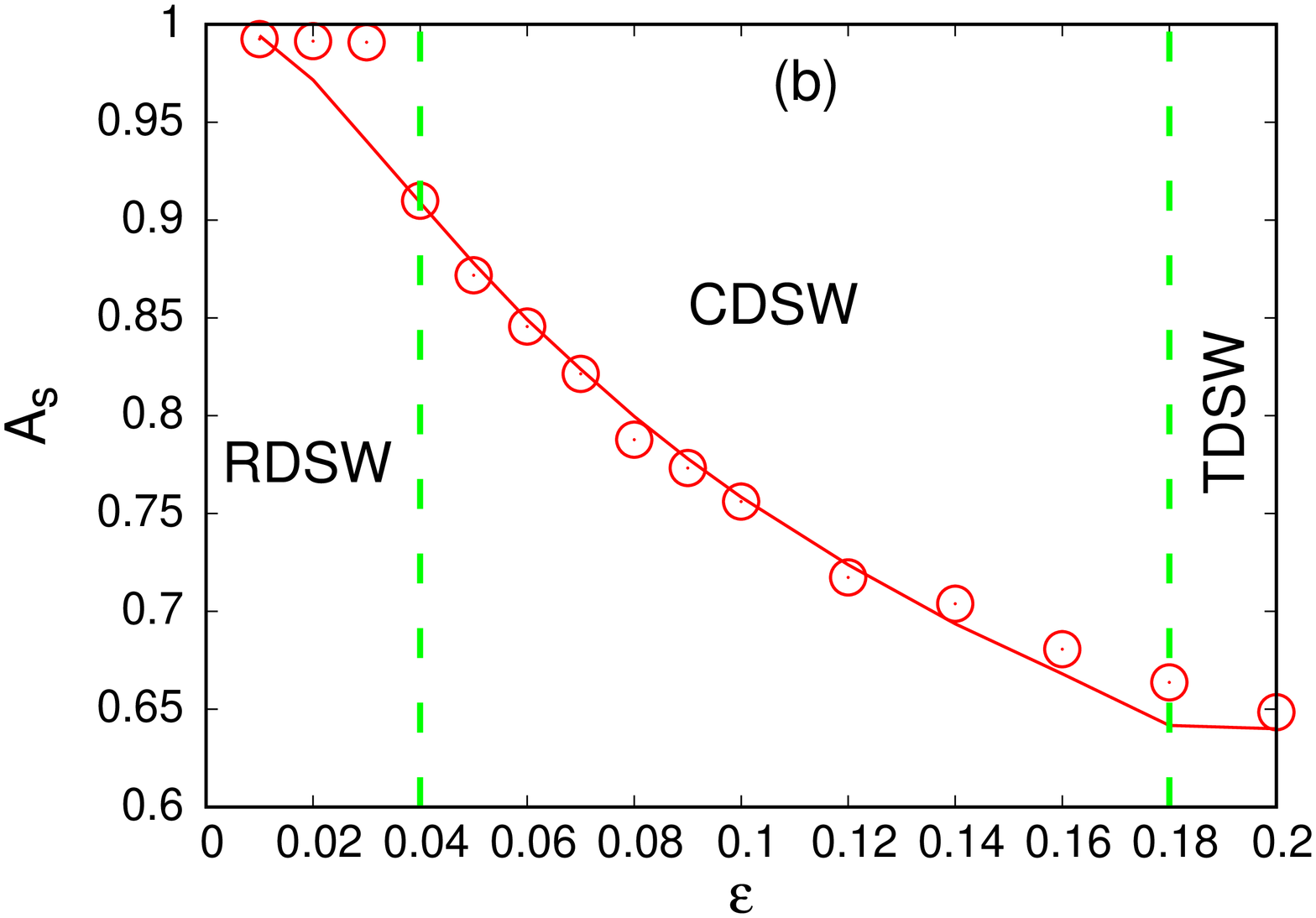}
    \includegraphics[angle=270,width=0.47\textwidth]{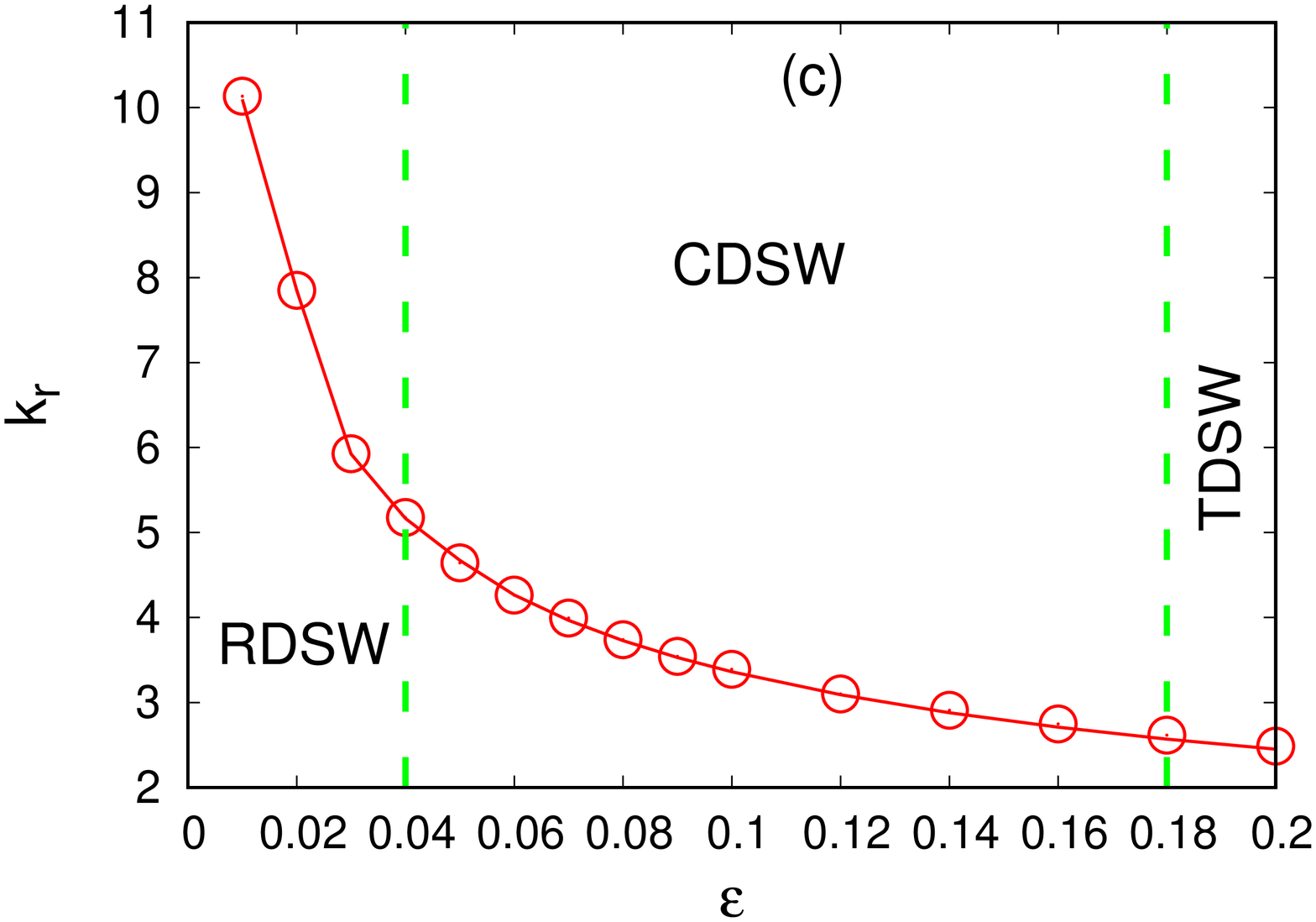}
    \includegraphics[angle=270,width=0.47\textwidth]{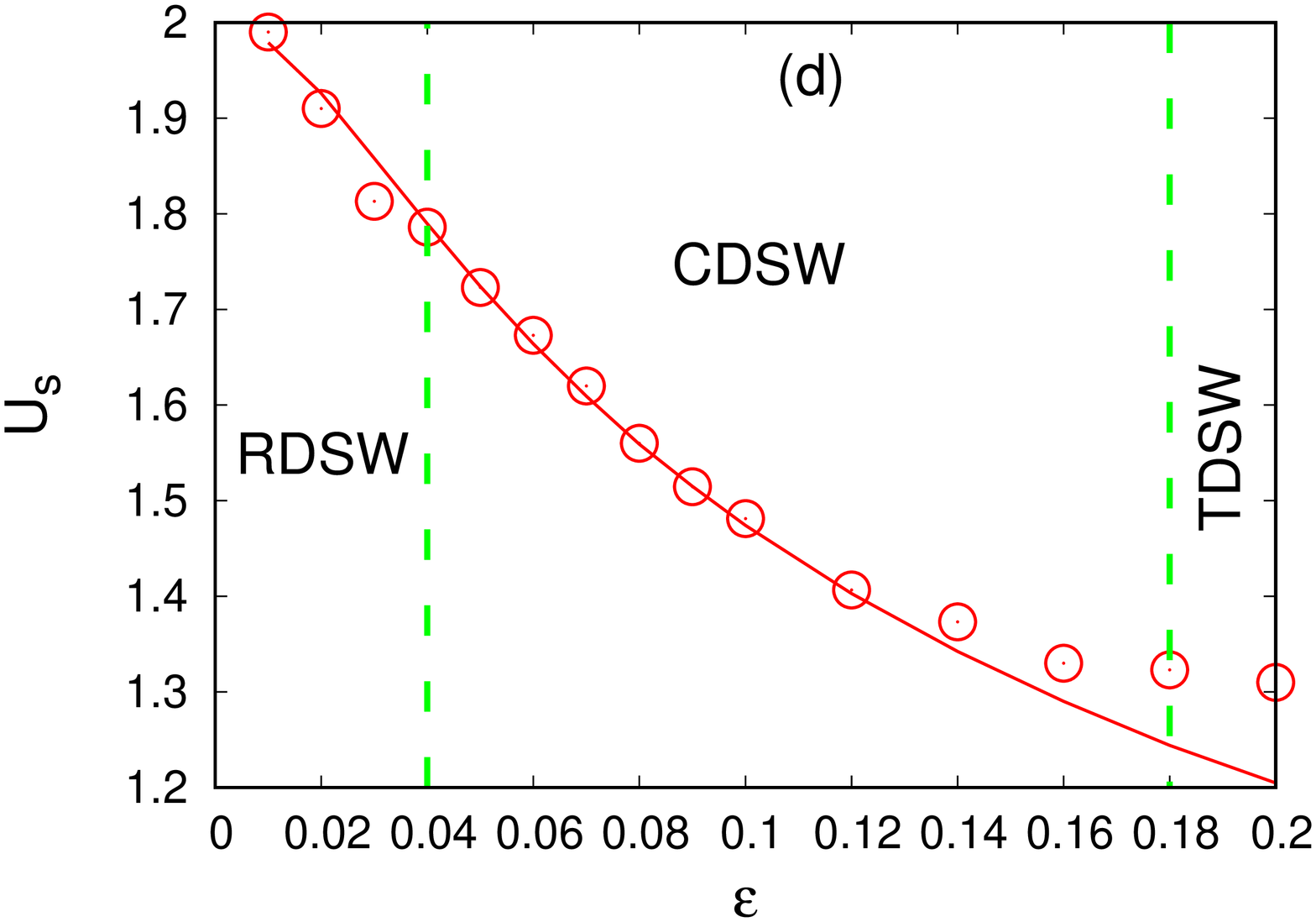}
    \caption{Comparison between numerical solutions of the Kawahara equation (\ref{e:kawahara}) and modulation theory.  Numerical solution:  \textcolor{red}{$\circ$} (red circle); modulation theory solution:  red (solid) line; boundaries between DSW regimes:  green (dashed) line. (a) resonant wave amplitude $A_{r}$, (b) solitary wave amplitude of CDSW $A_{s}$, (c) wavenumber of resonant wavetrain $k_{r}$, (d) velocity of Whitham shock $U_{s}$.  Here, $c_{4}=1$, $u_{-}=0.5$ and $u_{+}=0$.}
     \label{f:kawahara}
\end{figure}

The eKdV equation (\ref{e:ekdv}) was solved numerically using the pseudo-spectral method of Fornberg and Whitham \cite{bengt} as extended to enhance stability at high wavenumbers, particularly due to the higher order fifth order dispersion \cite{chan,tref}.  The spatial derivatives were calculated in Fourier space, with the equation propagated in time using the fourth order Runge-Kutta method in Fourier space.  As stated, to enhance
stability linear dispersion was propagated using an integrating factor \cite{chan,tref}.  Numerical solutions generated using this numerical scheme will now be compared with solutions of the modulation theory of Section \ref{s:modulation}.  As can be seen from Figure \ref{f:cdsw} that the resonant wavetrain is unstable.  The resonant wave amplitude was then calculated by averaging the amplitude over the resonant wavetrain up to its front. 

Figure \ref{f:kawahara} shows comparisons between full numerical solutions of the Kawahara equation (\ref{e:kawahara}) and modulation theory for the resonant wavetrain amplitude 
\begin{equation}
    A_{r} = a_{r} + a_{r}^{2}u_{2}, 
\end{equation}
see (\ref{e:stokes}), the CDSW solitary wave amplitude 
\begin{equation}
    A_{s}= a_{s}+\epsilon \left( c_6 + c_{7}\right) a^2_{s},
\end{equation}
see (\ref{eKdV_soliton}), the resonant wavetrain wavenumber $k_{r}$ and the Whitham shock velocity $U_{s}$.   In general, the agreement between modulation theory and numerical solutions is excellent across the CDSW regime.  The agreement even extends beyond the CDSW regime into the RDSW and TDSW regimes, particularly for the wavenumber of the resonant wavetrain $k_{r}$. We note that the resonant radiation is an unstable wavetrain (multi-phase wavetrain), so there exists no single dominant wavenumber. The numerical values of $k_{r}$ were then determined by averaging over the resonant wavetrain. It was found that averaging over 10 to 20 crests was sufficient. The agreement is good in the TDSW regime, which is expected as modulation theory in the TDSW regime is the limit of the present modulation theory if the amplitude $A_{s}$ of the solitary waves vanishes and the mean level becomes the level behind $u_{-}$ \cite{patjump}.  In this context, it should be noted that in the present work the solitary wave amplitude $A_{s}$ is measured from the level ahead $u_{+}=0$ as the equal amplitude approximation is based on the leading edge of the DSW, so that in the TDSW limit $A_{s} \to u_{-}$,
while in the work of Sprenger and Hoefer \cite{patjump} the wave amplitude is measured from the local mean level.  The agreement for the solitary wave amplitude $A_{s}$ is less good in the RDSW regime as the DSW cannot be approximated by a train of equal amplitude solitary waves since the DSW is a perturbed KdV DSW in this regime \cite{ekdv}. 

The present modulation theory gives that the amplitude $A_{r}$ of the resonant wavetrain rapidly approaches zero in the RDSW regime, as expected, but less rapidly than the numerical amplitude.  Similarly, the modulation and numerical Whitham shock velocities $U_{s}$ are in excellent agreement in the CDSW regime, and even in the RDSW regime.  The latter is expected as the RDSW solution is a perturbation of the generic DSW solution.  In the TDSW regime, the modulation Whitham shock velocity differs slightly from the numerical velocity.

Table \ref{t:kawmeanu} displays the modulation theory solution for the mean level $\bar{u}_{s}$ of the Kawahara CDSW for a range of $\epsilon$.  As stated above, the Whitham shock results in a very small change in mean level from that ahead, $u_{+}$, which validates the assumption above used to solve the system of modulation equations for the CDSW.  

Table \ref{t:meansolekdv} shows the modulation solution for the mean height $\bar{u}_{s}$ of the CDSW solitary waves for a general eKdV equation.  As for the Kawahara equation results of Table \ref{t:kawmeanu}, the jump in mean height across the Whitham shock is minimal, so much so that it is difficult to measure from numerical solutions.  
We then conclude that the resonant wavetrain is again essentially linear.  As the mean level and the wave propagating on this mean level do not couple at the linear level \cite{whitham}, the 
minimal variation of $\bar{u}_{r}$ from $u_{+}$, as assumed in order to solve the modulation equations, is 
then clear.

\begin{figure}[!ht]
    \centering
    \includegraphics[angle=270,width=0.47\textwidth]{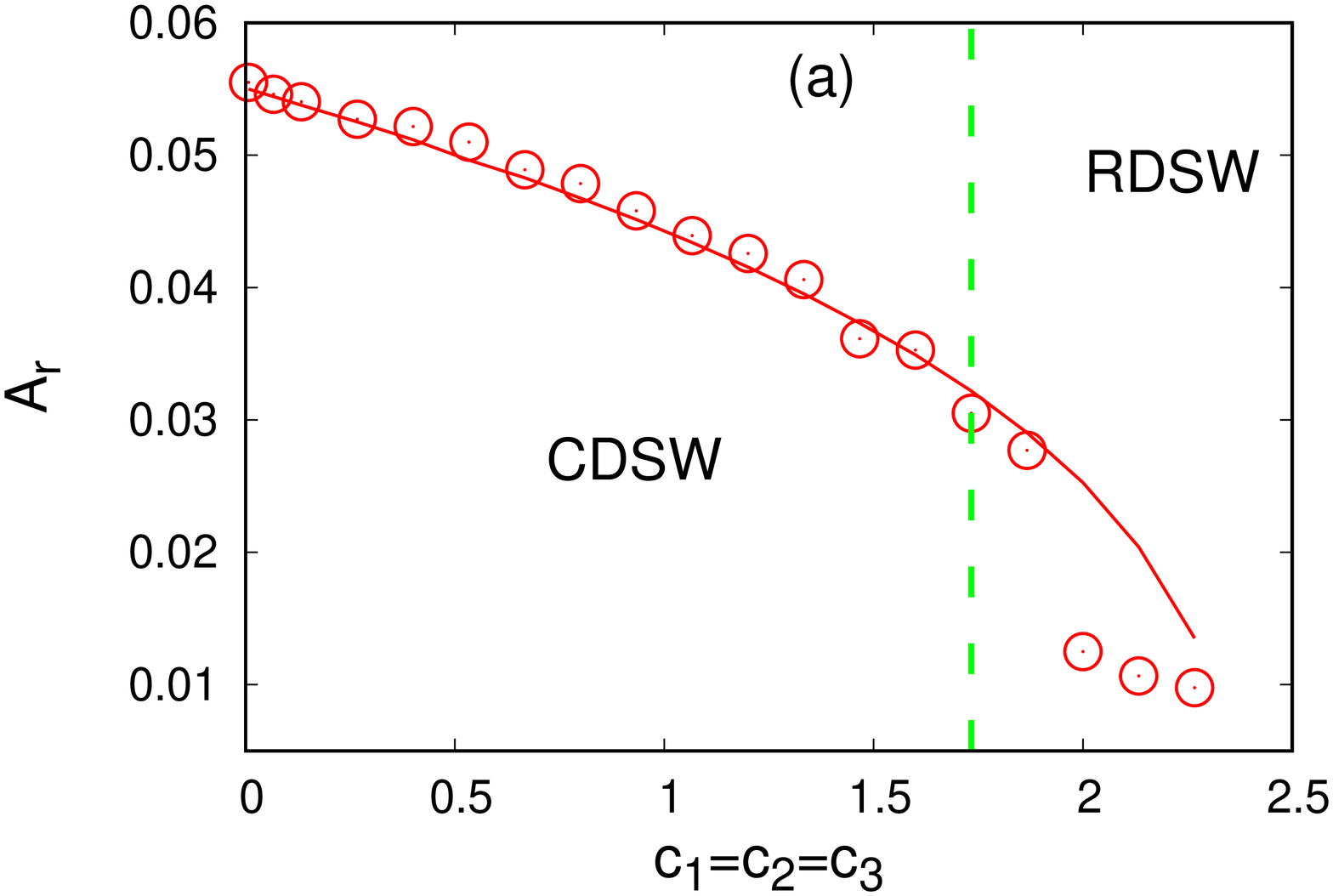}
    \includegraphics[angle=270,width=0.47\textwidth]{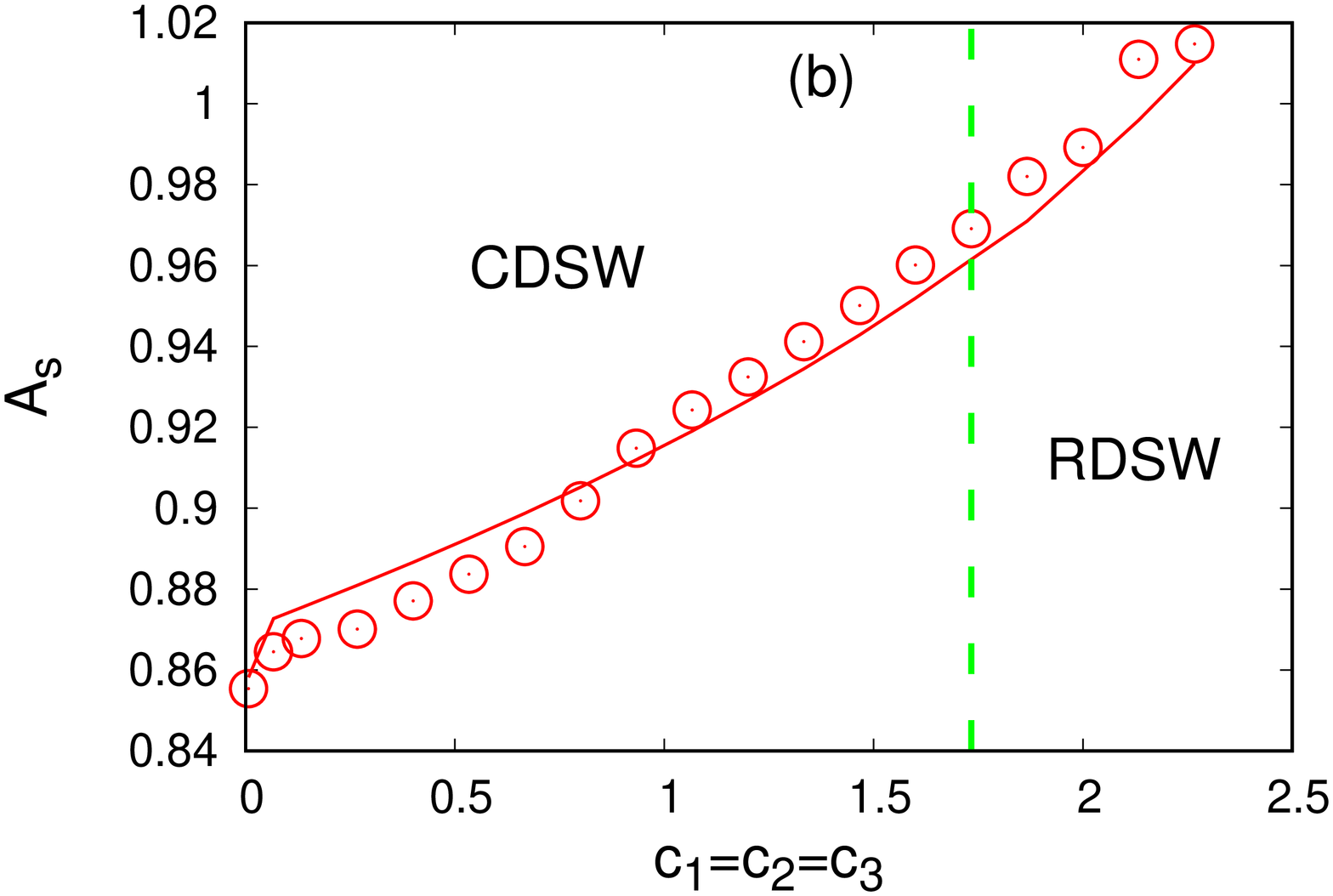}
    \includegraphics[angle=270,width=0.47\textwidth]{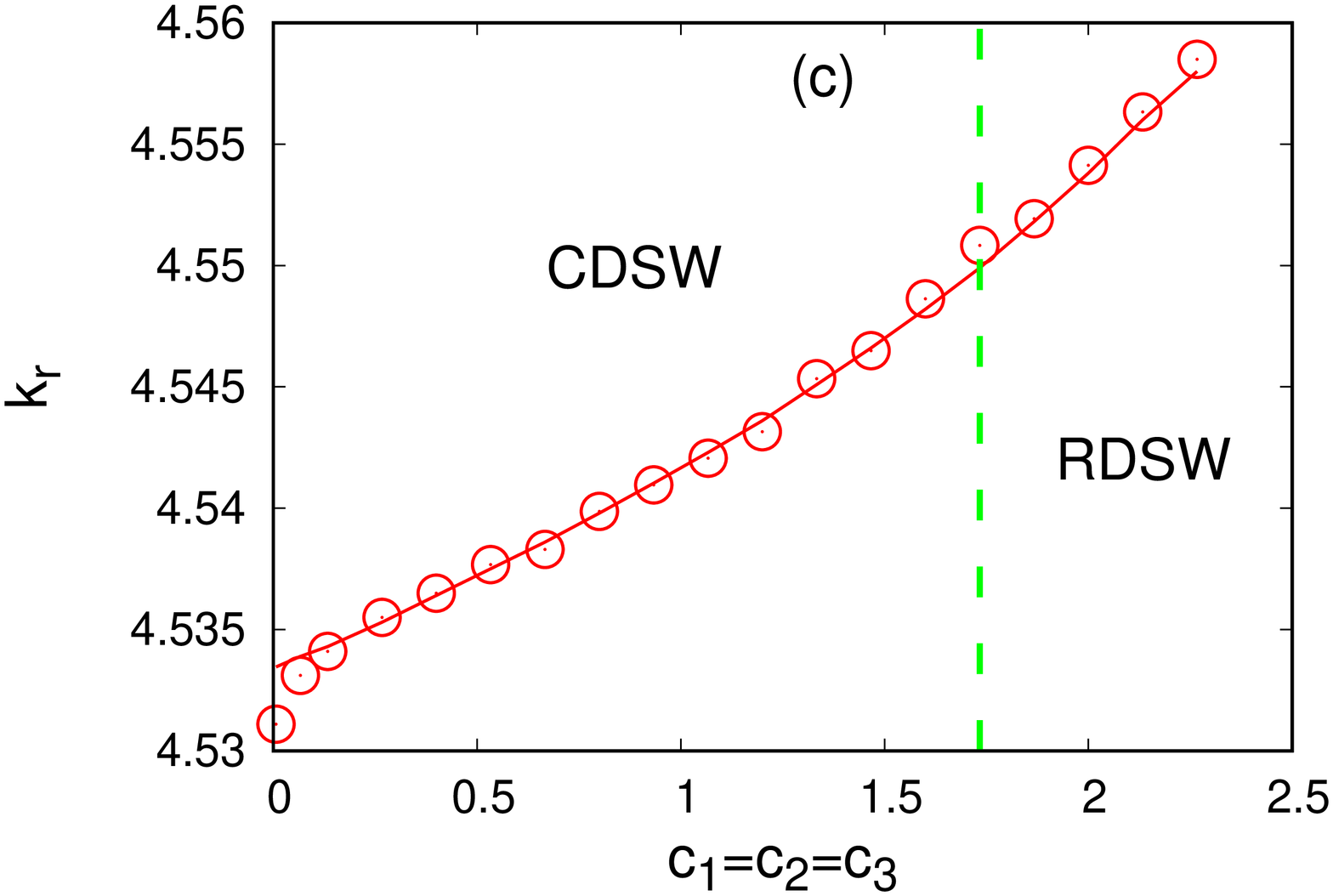}
    \includegraphics[angle=270,width=0.47\textwidth]{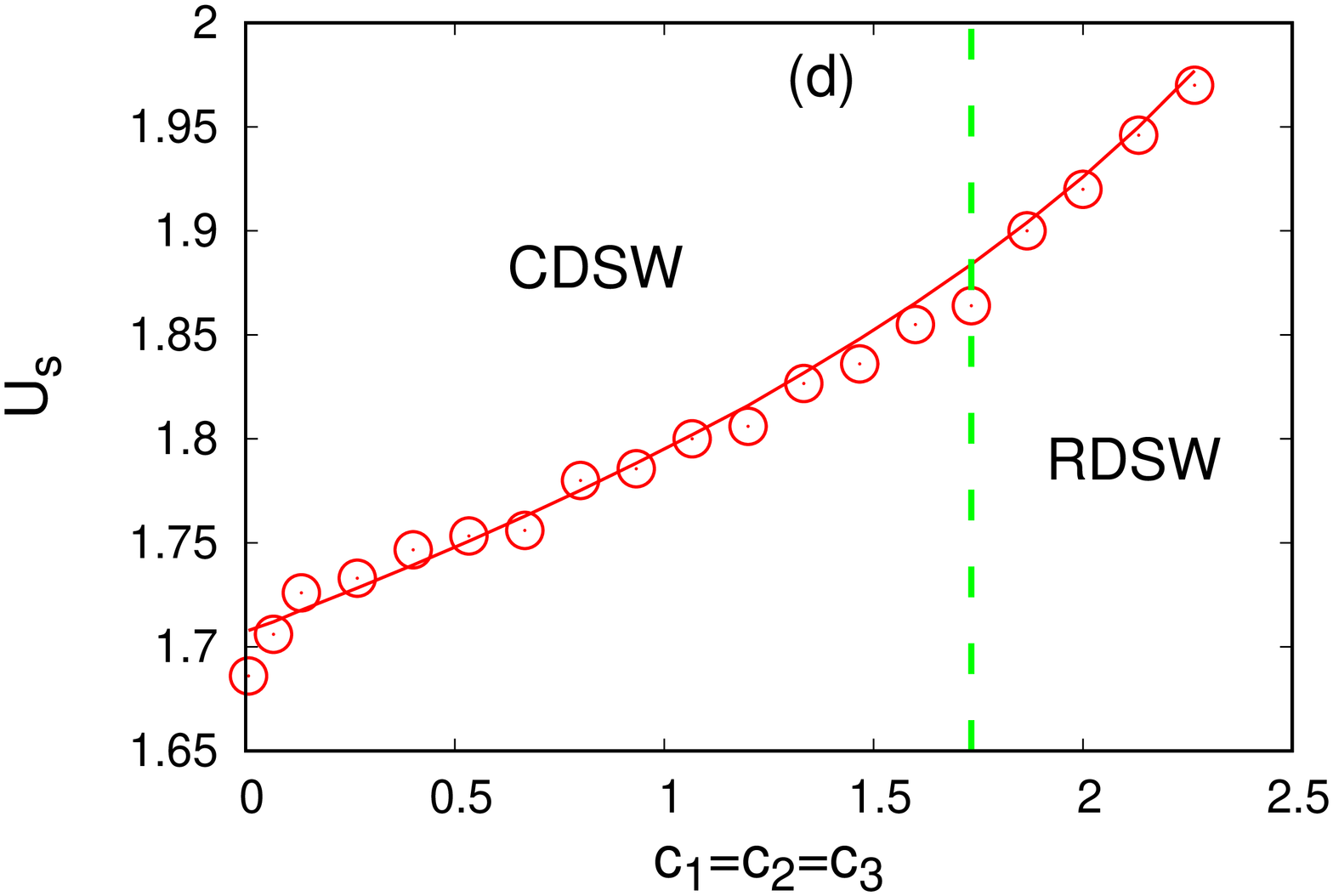}
    \caption{Comparison between numerical solutions of the eKdV equation (\ref{e:ekdv}) with $c_{1}=c_{2}=c_{3}$ and $c_{4}=0.351$ with modulation theory.  Numerical solution:  \textcolor{red}{$\circ$} (red circle); modulation theory solution:  red (solid) line; boundaries between DSW regimes:  green (dashed) line.  (a) resonant wave amplitude $A_{r}$, (b) solitary wave amplitude of CDSW $A_{s}$, (c) wavenumber of resonant wavetrain $k_{r}$, (d) velocity of Whitham shock $U_{s}$.   Here, $\varepsilon=0.15$, $u_{-}=0.5$ and $u_{+}=0$.}
    \label{f:ekdv}
\end{figure}

Figure \ref{f:ekdv} shows similar comparisons as Figure \ref{f:kawahara} for the Kawahara equation, but for a general eKdV equation with $c_{1}$, $c_{2}$ and $c_{3}$ non-zero.  The overall agreement is similar to that for the Kawahara equation in the CDSW regime. 
The modulation theory resonant wave amplitude $A_{r}$ 
is in excellent agreement with the numerical amplitude, with reasonable agreement even in the RDSW regime, as for the Kawahara equation.  In contrast, the modulation solitary wave amplitude $A_{s}$ is in excellent agreement with the numerical amplitude, even in the RDSW regime, for which good agreement is not expected, as discussed above for the Kawahara equation results.  Figures \ref{f:ekdv}(c) and (d) show similar excellent agreement for the resonant wavetrain wavenumber $k_{r}$ and the Whitham shock velocity $U_{s}$, with the excellent agreement holding into the RDSW regime.  The Whitham shock velocity $U_{s}$ and the resonant wavenumber $k_{r}$ are connected through the resonance condition (\ref{e:resonance}), so this similar agreement with numerical solutions is expected.

To additionally ensure the validity of the equal amplitude approximation used for the unstable bore of the CDSW, we report a few comparison cases for the number of leading, randomly distributed solitary waves $N$, on neglecting the descending waves near the trailing edge which take the solution down to the initial level behind $u_{-}$, as compared with numerical results. In theory, $N$ is given by equation (\ref{e:masscdsw}) or equation (\ref{e:energycdsw}). For example, with $\epsilon=0.04$, $c_{4}=1$ ($\epsilon=0.14$, $c_{4}=1$) at $t=30$, Kawahara CDSW modulation theory gives $N=8.6580\approx{9}$ ($N=10.16\approx{10}$) and numerical solutions give $N\approx{9}$ ($N\approx{11}$). Similarly, with $\epsilon=0.15$, $c_{1}=c_{2}=c_{3}=0.006$ ($c_{1}=c_{2}=c_{3}=1.333$), $c_{4}=0.3513$ at $t=30$, eKdV CDSW modulation theory gives $N=8.9041\approx{9}$  ($N=8.463\approx{8}$), and numerical solutions give $N\approx{10}$ ($N\approx{8}$).

\begin{table}
\centering
\begin{tabular}{|l|l|l|l|l|} 
\hline
 $\epsilon=0.01$ & $\epsilon=0.03$ & $\epsilon=0.05$ &  $\epsilon=0.1$ & $\epsilon=0.2$ \\ 
\hline
 $\bar{u}_{s}=5.03\times{10^{-6}}$ & $\bar{u}_{s}=9.6\times{10^{-5}}$ & $\bar{u}_{s}=3.05\times{10^{-4}}$ & $\bar{u}_{s}=1.1\times{10^{-3}}$ &  $\bar{u}_{s}=3.20\times{10^{-3}}$  \\
\hline
\end{tabular}
\caption{Modulation mean level $\bar{u}_{s}$ of CDSW for Kawahara equation (\ref{e:kawahara}).  Here, $c_{4}=1$, $u_{-}=0.5$ and $u_{+}=0$.}
\label{t:kawmeanu}
\end{table}

\begin{table}
\centering
\begin{tabular}{|l|l|l|l|l|} 
\hline
 $c_{1} = 0.006$ & $c_{1} = 0.660$ & $c_{1} = 1.333$ &  $c_{1} = 2.000$ & $c_{1} = 2.266$ \\ 
\hline
 $\bar{u}_{s}=3.4\times{10^{-4}}$ & $\bar{u}_{s}=5.3\times{10^{-4}}$ & $\bar{u}_{s}=9.3\times{10^{-4}}$ & $\bar{u}_{s}=1.58\times{10^{-3}}$ &  $\bar{u}_{s}=1.96\times{10^{-3}}$  \\
\hline
\end{tabular}
\caption{Modulation mean level $\bar{u}_{s}$ of the CDSW for the eKdV equation (\ref{e:ekdv}).  The coefficients $c_{1}$, $c_{2}$ and $c_{3}$ are taken equal, while $c_{4}$ is fixed at $c_{4}=0.351$.  Here, $\epsilon=0.15$, $u_{-}=0.5$ and $u_{+}=0$.}
\label{t:meansolekdv}
\end{table}

\section{Water Waves}

\begin{figure}
    \centering
    \includegraphics[angle=270,width=0.47\textwidth]{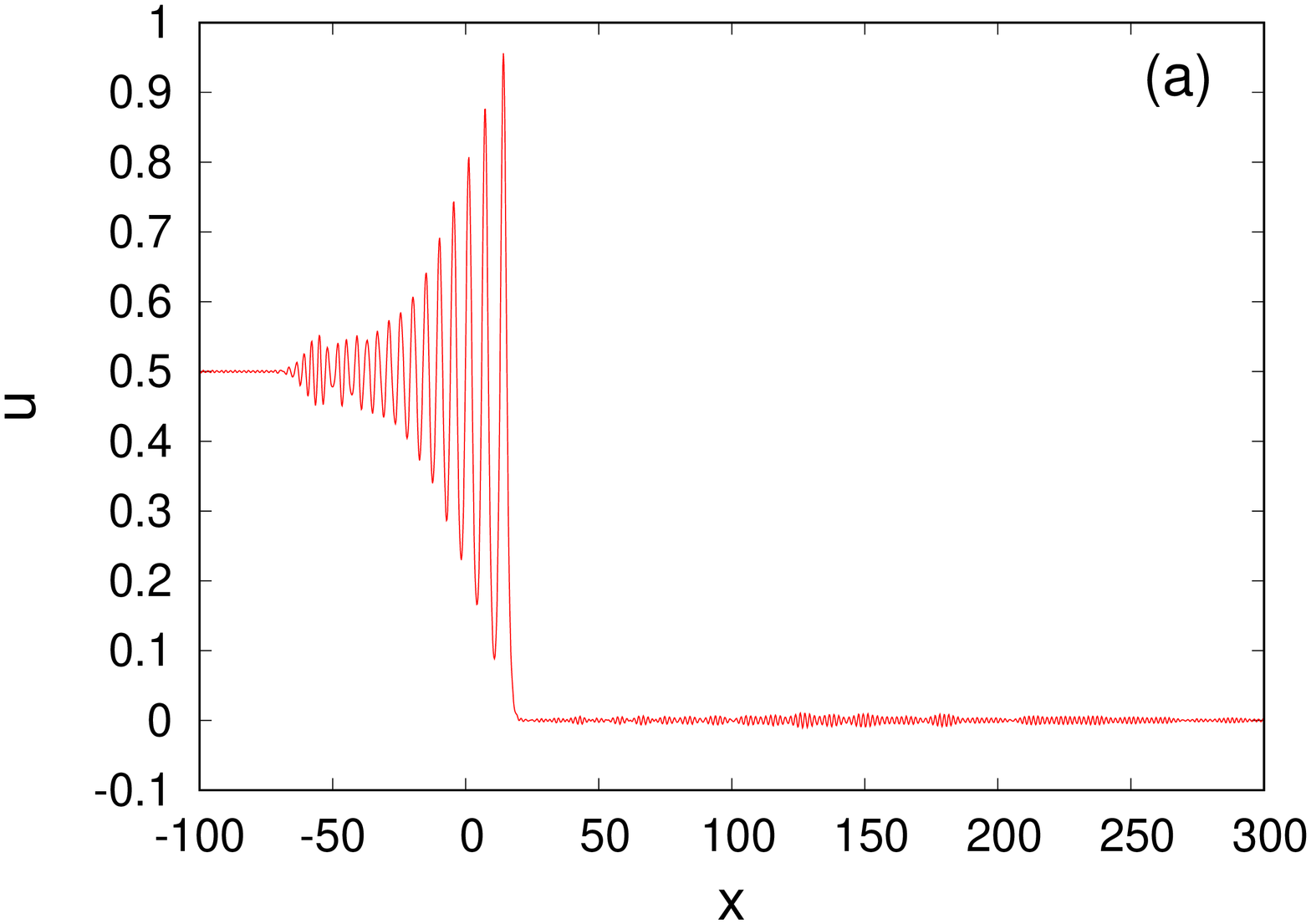}
    \includegraphics[angle=270,width=0.47\textwidth]{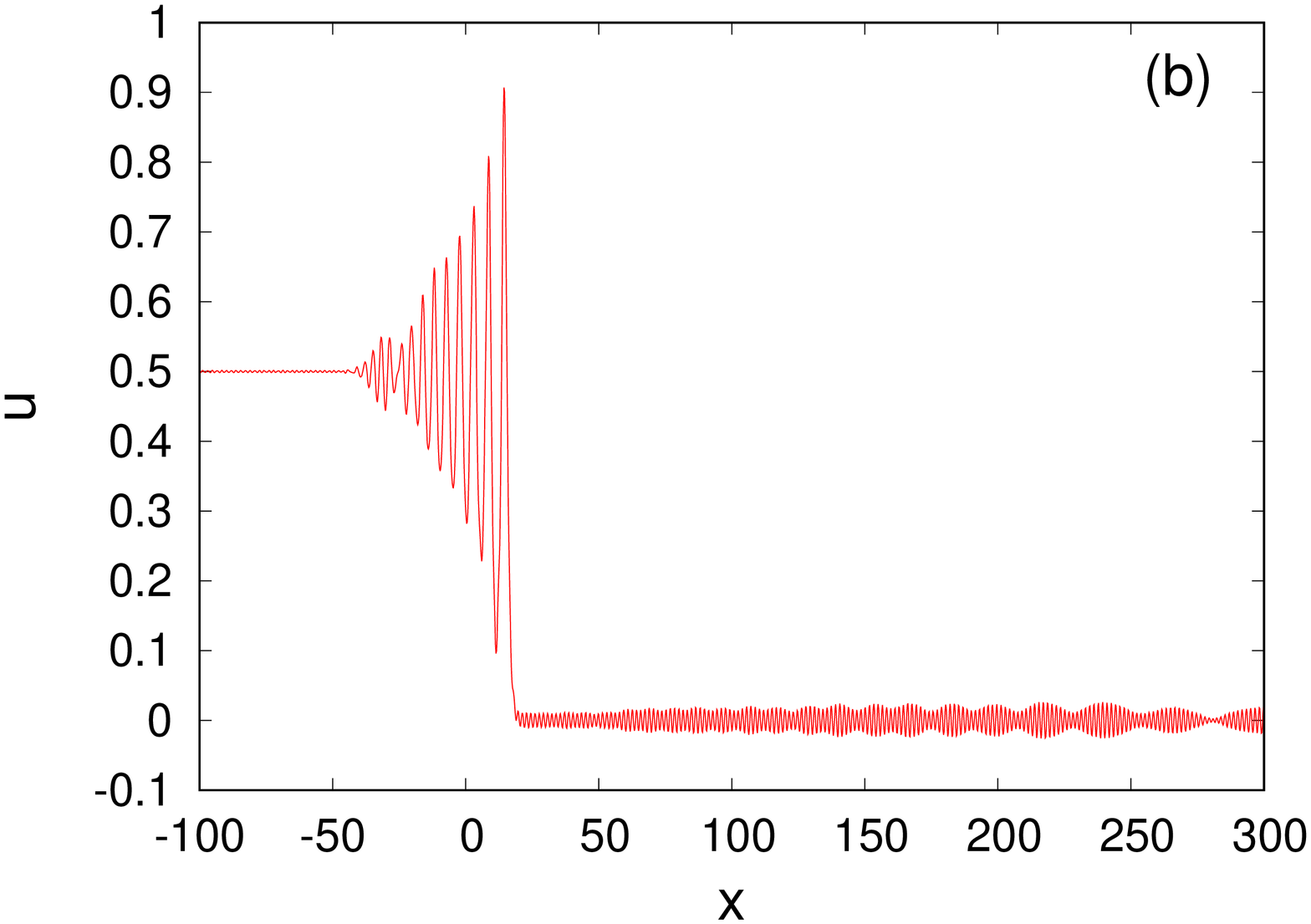}
    \includegraphics[angle=270,width=0.47\textwidth]{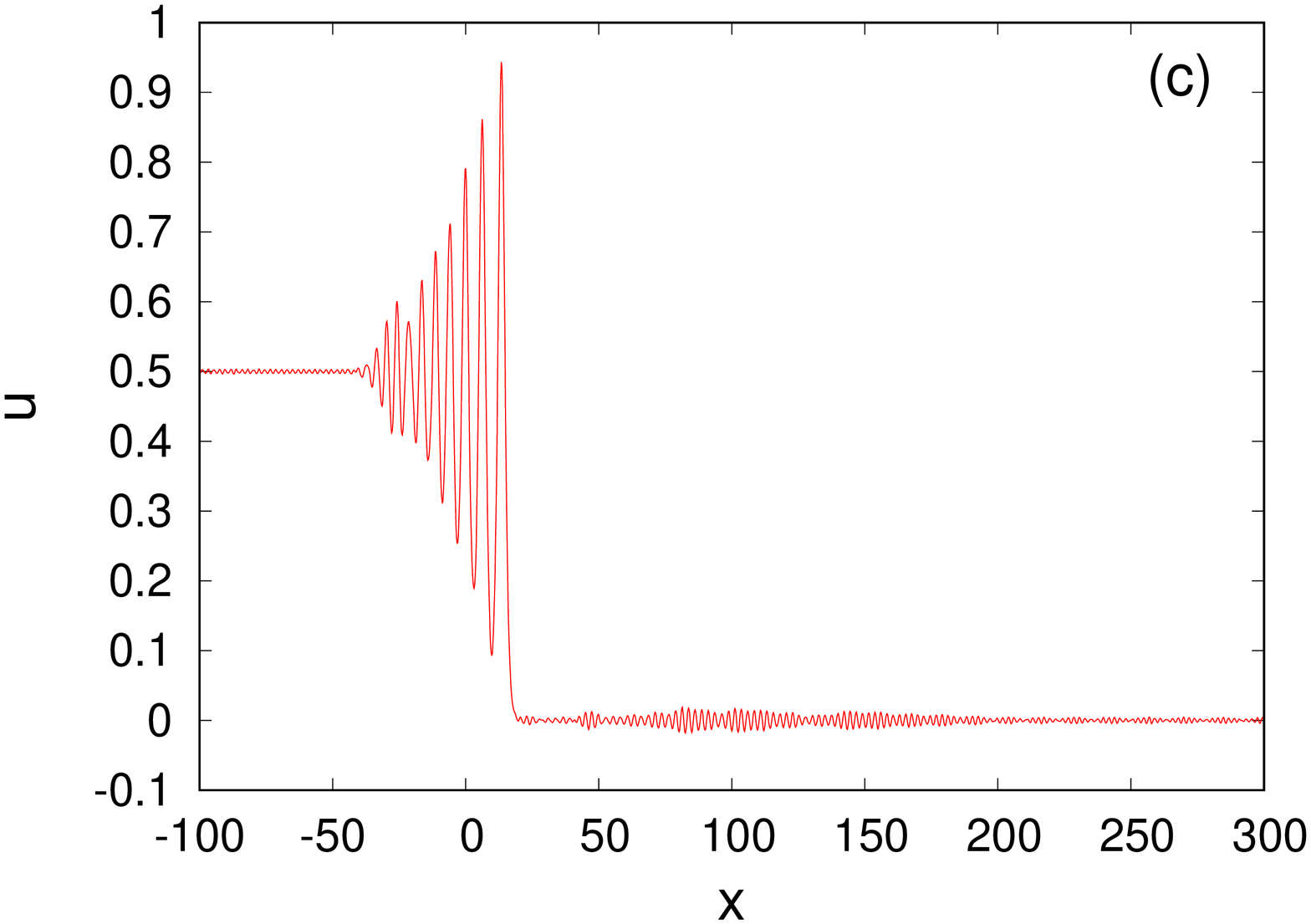}
    \includegraphics[angle=270,width=0.47\textwidth]{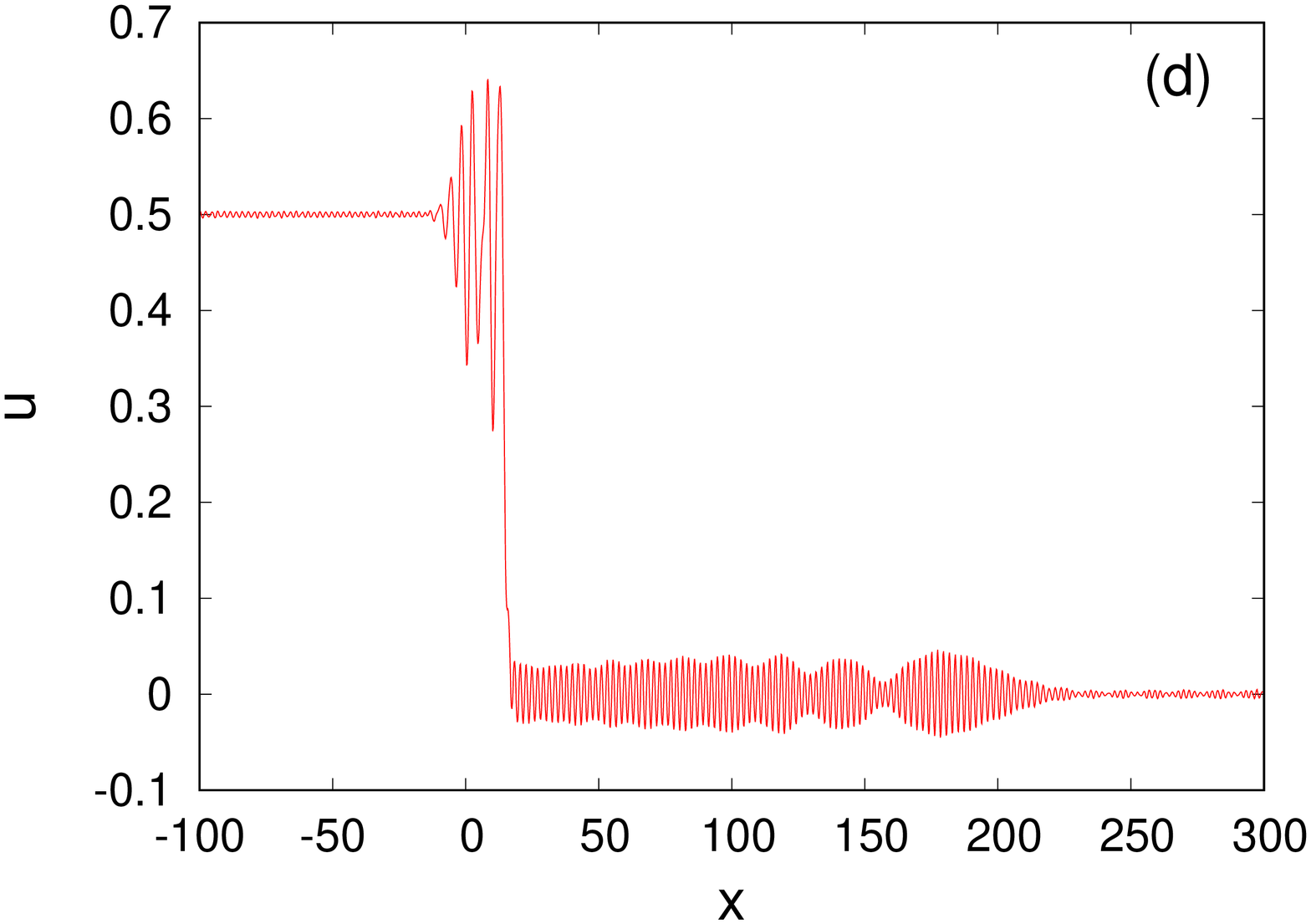}
    \caption{Dependence of bore structure on higher order parameters with $c_{1} = -3/2$, $c_{2} = 23/4$ and $c_{4} = 19/40$. (a) $c_{3}=5/2$, $\epsilon =0.15$, (b) $c_{3}=0$, $\epsilon=0.15$, (c) $c_{3}=5/2$, $\epsilon=0.3$, (d) $c_{3}=0$, $\epsilon=0.3$.  Here, $u_{-}=0.5$, $u_{+}=0$ and $t=10$.}
     \label{f:waterwave}
\end{figure}

One of the motivations behind the present work is the 
observed resonant wave amplitude minimum for the eKdV equation (\ref{e:ekdv}) with the water wave coefficients \cite{resekdv}.  This is illustrated in Figure \ref{f:waterwave} based on two values of the 
nonlinearity parameter $\epsilon$. 
Figure \ref{f:waterwave}(a) displays the water wave bore for $\epsilon=0.15$.  The bore is in the RDSW regime and the resonant wavetrain has amplitude $\sim 6 \times 10^{-3}$.  To illustrate the effect of the higher order coefficients on the bore structure, Figure \ref{f:waterwave}(b) displays the bore solution for the water wave coefficients $c_{1}$, $c_{2}$ and $c_{4}$, but with $c_{3}=0$. 
The coefficient $c_{3}$ was varied as this coefficient was found to have the greatest effect on the resonant wave amplitude.  The bore has become unstable and is bordering on the CDSW regime with the resonant wavetrain having amplitude $\sim 2 \times 10^{-2}$.  The increase of $\epsilon$ to $0.3$ shows the same overall behaviour.  The bore with the water wave coefficients, Figure \ref{f:waterwave}(c), is bordering on the CDSW regime, with the resonant wavetrain still having minimal amplitude $\sim 1 \times 10^{-2}$.  In contrast with $c_{3}=0$, Figure \ref{f:waterwave}(d), the bore is bordering on the TDSW regime, with the waves of the bore having much reduced amplitude and extent, and the resonant wavetrain having amplitude $\sim 4 \times 10^{-2}$.  As well as greatly reducing the resonant wave amplitude, the water wave coefficients delay the onset of the transition between the bore regimes, RDSW to CDSW to TDSW.  As the resonant wavetrain amplitude increases on transition from RDSW to CDSW to TDSW, these two effects are connected.  The modulation theory of the present work will now be used to analyse the effect of the values of $c_{i}$, $i=1,\ldots, 4$, on the bore structure.  

\begin{figure}[!ht]
\centering
\includegraphics[angle=270,width=0.47\textwidth]{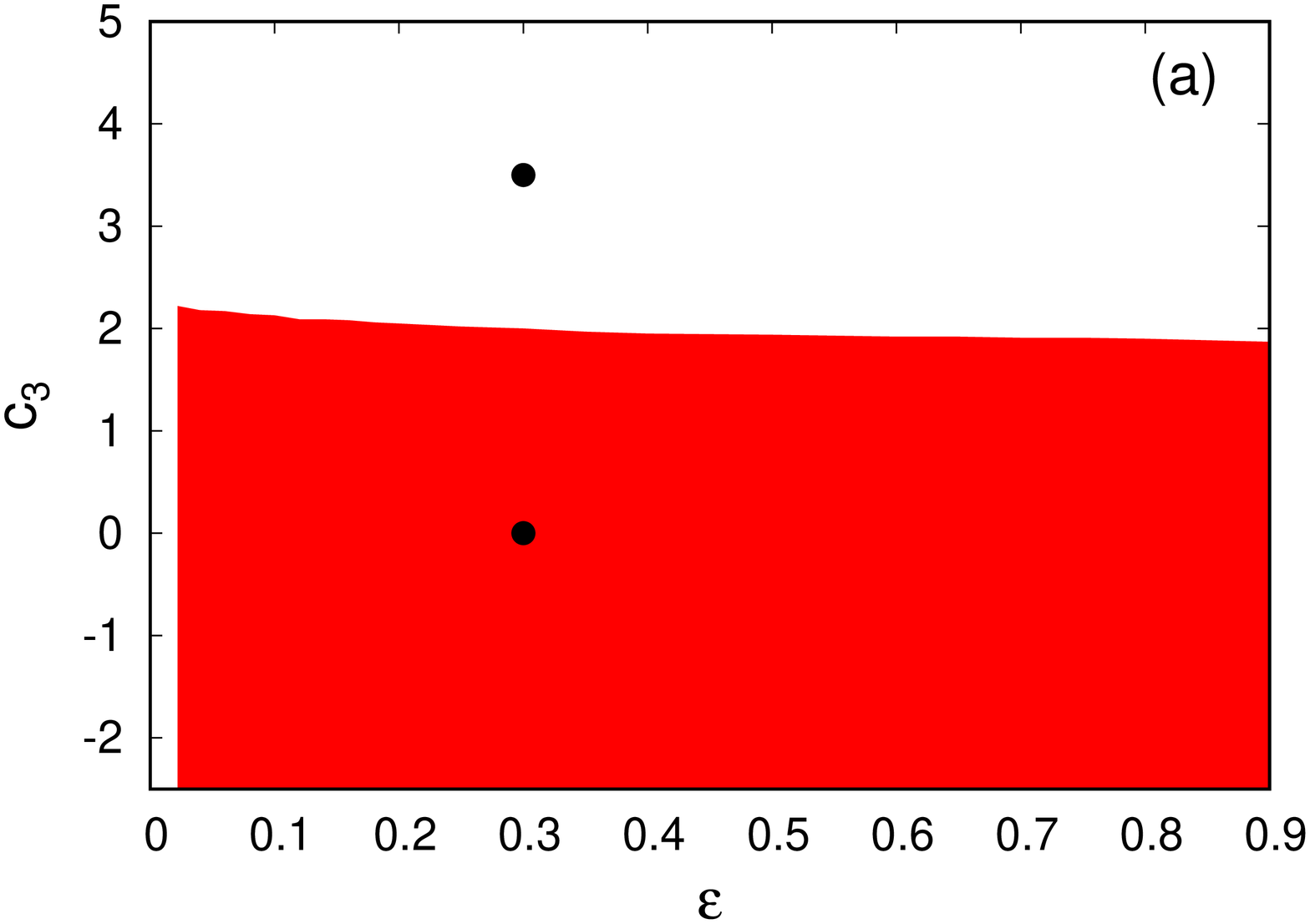}
    \includegraphics[angle=270,width=0.47\textwidth]{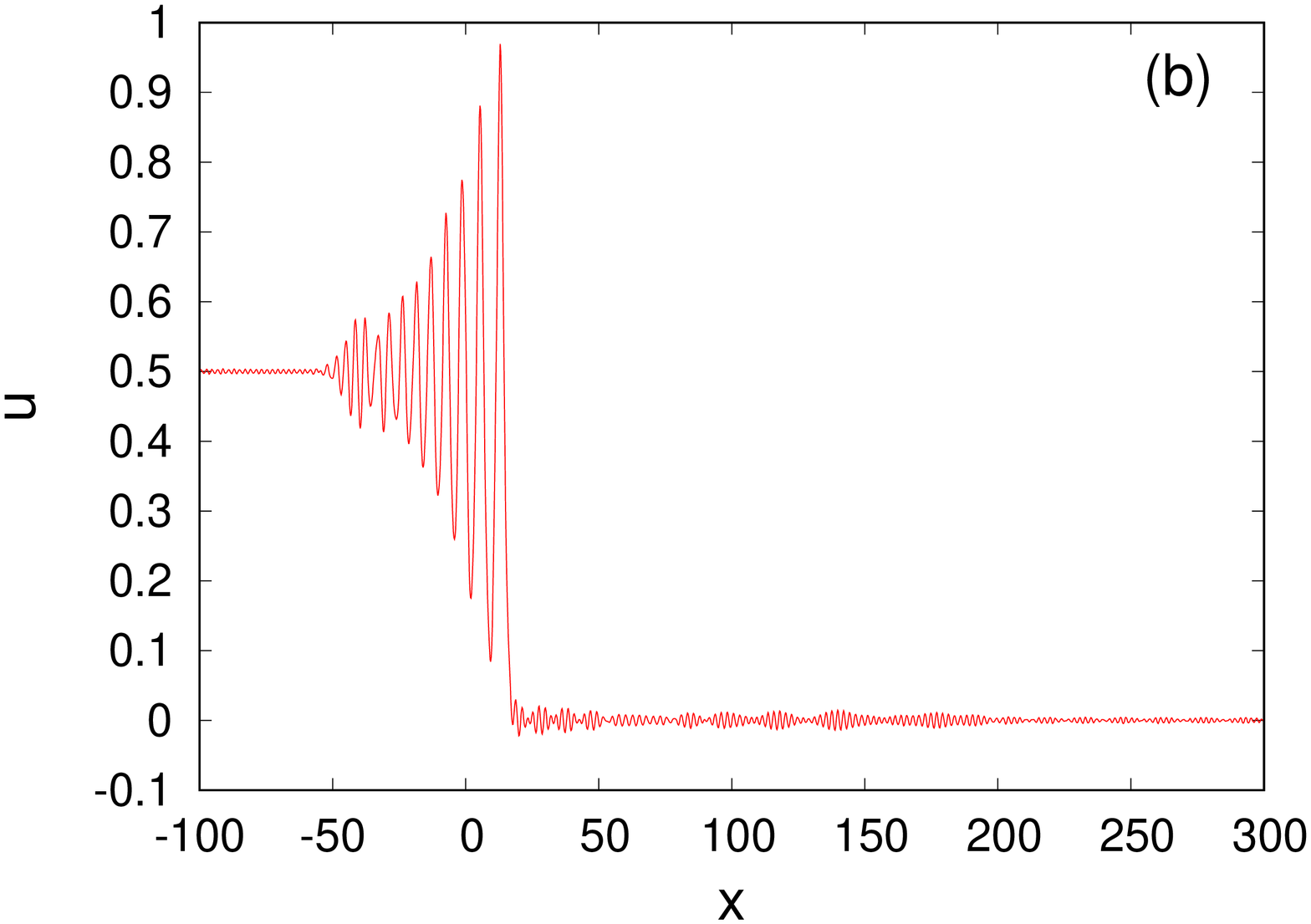}
\caption{(a) Existence interval for the modulation solution as the nonlinearity $\epsilon$ and $c_{3}$ vary.  The modulation solution exists in the red (shaded) region.  (b) Example of bore for $c_{3}=3.5$ and $\epsilon = 0.3$ at $t=10$. The other coefficients are the water wave values $c_{1}=-3/2$, $c_{2}=23/4$ and $c_{4}=19/40$. Here, $u_{-}=0.5$ and $u_{+}=0$.}
\label{f:c3}
\end{figure}

The modulation equations with the Whitham shock jump conditions for the eKdV equation do not have a (real) solution when the coefficients $c_{i}$, $i=1,\ldots,4$,
take the water wave values.  However, a solution of the modulation equations does exist for ranges of the higher order coefficients.  To be specific, the dependence of the existence of modulation theory solutions on the higher order coefficients will be explored by varying the nonlinearity $\epsilon$ with $c_{1}$, $c_{2}$ and $c_{4}$ taking the water wave values and determining the existence ranges in the final higher order coefficient $c_{3}$.  Figure \ref{f:c3}(a) shows the existence range of the modulation theory solution for a range of the nonlinearity $\epsilon$, up to high values which are outside the range of asymptotic validity of the eKdV equation, noting that the water wave value is $c_{3}=5/2$.  Figure \ref{f:c3}(b) shows a bore for $c_{3}=3.5$ and $\epsilon = 0.3$, which is in the region for which the modulation equations do not have a solution.  The location of this $c_{3}$ value is shown by the upper black dot in Figure \ref{f:c3}(a).   Comparing with Figure \ref{f:waterwave}(d) it can be seen that the resonant wavetrain has greatly reduced amplitude, in agreement with modulation theory.  The location of the $c_{3}$ value used for Figure \ref{f:waterwave}(d) is shown by the lower black dot in Figure \ref{f:c3}(a).   The resonant wave amplitude of Figure \ref{f:c3}(b) is of the same order as the example shown in Figure \ref{f:waterwave}(c), which is for the water wave coefficients for the same value of $\epsilon$.  While this is not conclusive justification for the low resonant wave amplitudes seen in Figures \ref{f:waterwave}(a) and (c) and \ref{f:c3}(b) as the bore for $\epsilon = 0.15$ is in the RDSW regime and for $\epsilon = 0.3$ is just in the CDSW regime and the present modulation theory is for the CDSW regime, it is consistent with the observed numerical results
and provides justification to some degree.  A final observation from these modulation theory results is that the minimal resonant wave amplitude exists for an unbounded range of $c_{3}$.  This is in contrast 
to the conclusion of \cite{resekdv} that an amplitude node exists for a discrete combination of the higher order coefficients $c_{i}$, $i=1,\ldots,4$, with small amplitude in a neighbourhood of this node.  This conclusion was based on previous results for resonant solitary waves governed by the eKdV equation, as detailed in this work.
It can then be concluded that results for solitary waves do not necessarily transfer to bores governed by the same equation.

\begin{figure}[!ht]
\centering
\includegraphics[angle=270,width=0.47\textwidth]{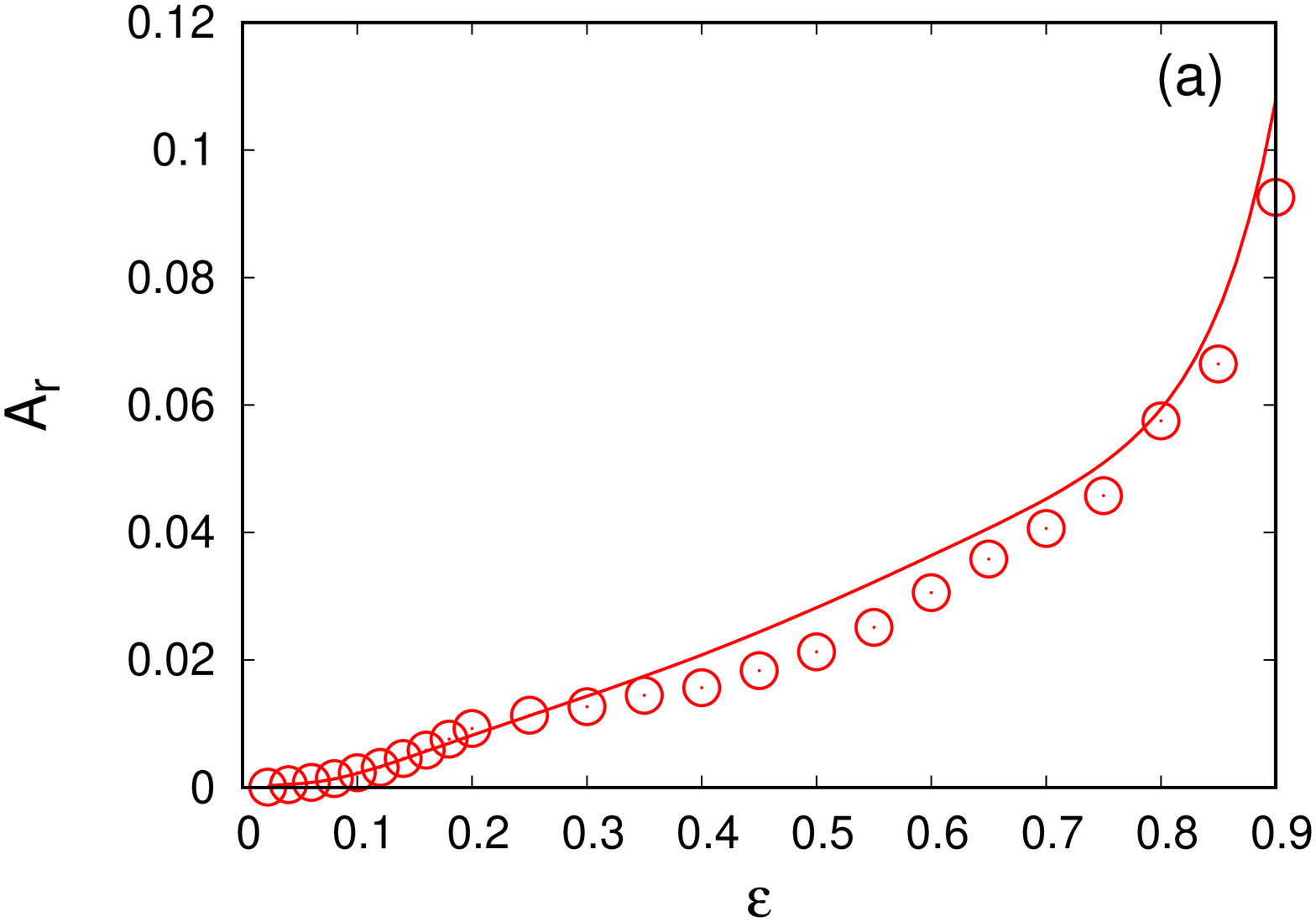}
\includegraphics[angle=270,width=0.501\textwidth]{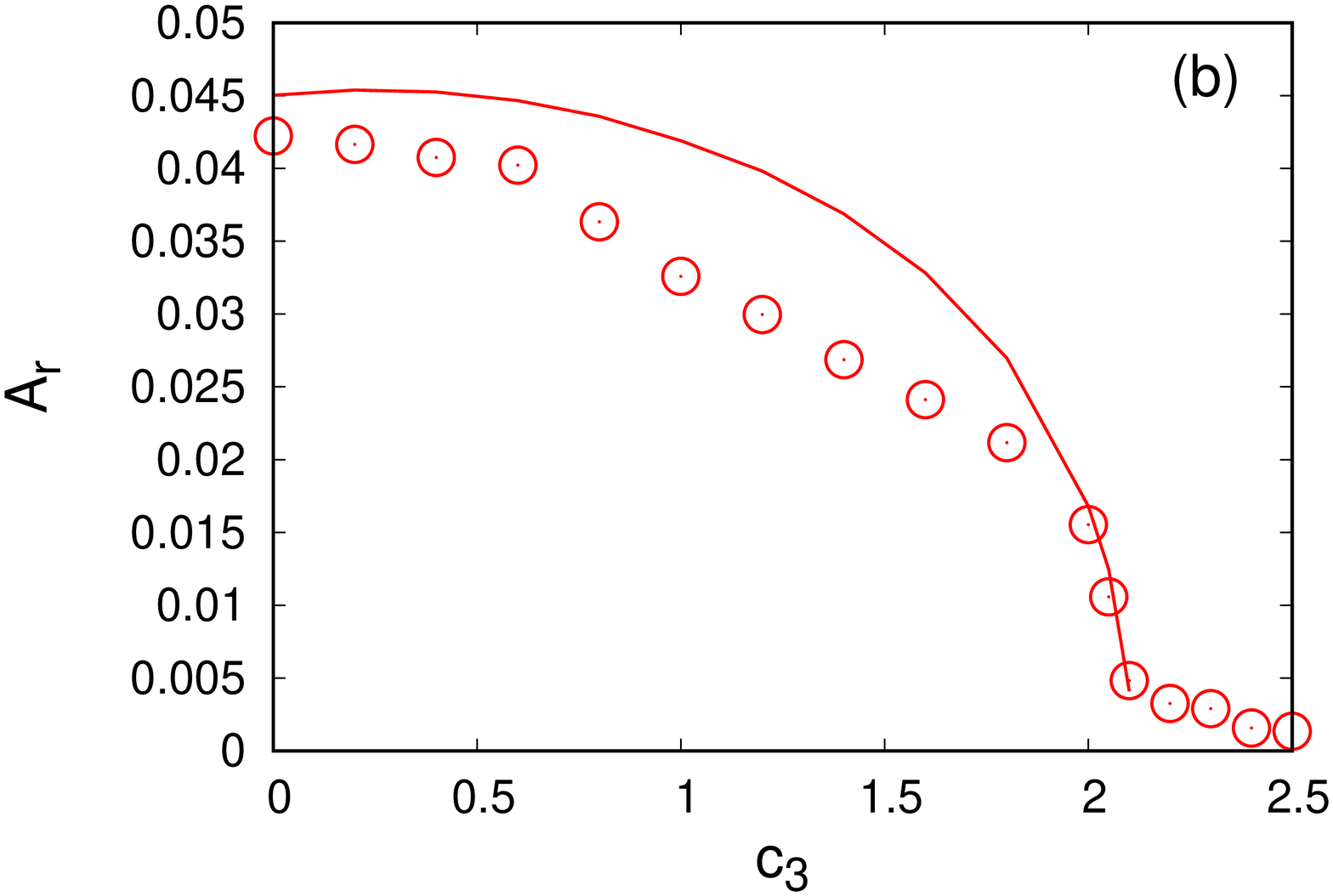}
\caption{(a) Resonant wave amplitude $A_{r}$ just below the existence borderline of Figure \ref{f:c3}. (b) Resonant wave amplitude $A_{r}$ as $c_{3}$ varies with $\epsilon = 0.13$.  Modulation theory amplitude:  red (solid) line; numerical amplitude:  $\textcolor{red}{\circ}$ red circle.   The other coefficients are the water wave values $c_{1}=-3/2$, $c_{2}=23/4$ and $c_{4}=19/40$. Here, $u_{-}=0.5$ and $u_{+}=0$.}
\label{f:radamp}
\end{figure}

Figure \ref{f:radamp}(a) shows the resonant wave amplitude $A_{r}$ just below the modulation theory cutoff of Figure \ref{f:c3}(a) as given by modulation theory and full numerical solutions.  
It can be seen that the agreement between theory and numerical solutions is excellent and that the 
resonant wave amplitude is small, even up to very large values of the nonlinearity parameter $\epsilon$.  
Figure \ref{f:radamp}(b) details the rapid decrease
of the resonant wave amplitude $A_{r}$ as the modulation theory borderline of Figure \ref{f:c3}(a) is crossed.  For this figure the higher order coefficient $c_{3}$ was varied, while $c_{1}$, $c_{2}$ and $c_{4}$ were kept at their water wave values.  The nonlinearity parameter used was $\epsilon = 0.13$, for which modulation theory gives the borderline $c_{3} = 2.1$.  The rapid decrease of both the numerical and modulation theory amplitudes as the theoretical borderline is approached as $c_{3}$ increases is clearly visible, with the numerical amplitude being very small above $c_{3} = 2.1$.  Indeed, as $c_{3}$ increases above $2.1$, the numerical amplitude continues to decrease.  Finally, the agreement between the numerical and modulation theory resonant wave amplitudes is good up to the cut-off.

\section{Conclusions}

Whitham modulation theory has been developed to obtain the cross-over dispersive shock wave (CDSW) solution of the extended Korteweg-de Vries equation.  The DSW itself in this regime is unstable and has a different structure to the standard Korteweg-de Vries DSW, consisting of a train of solitary waves of equal amplitude on average, instead of a modulated train of waves of nearly linearly decreasing amplitude from leading to trailing edges.  This non-standard structure has been exploited to obtain an approximate solution for the DSW.  The resonant wavetrain was obtained as a Stokes wave.  The key concept of a Whitham shock, a jump in the modulated parameters of a wavetrain, was used to link the CDSW and the resonant wavetrain.  It was found that this combination of modulation theory and approximate theory gave solutions in excellent agreement with full numerical solutions of the eKdV equation.  

The modulation theory developed in the present work
was used to successfully explain the numerically observed minimal resonant wave amplitude when the higher order coefficients in the eKdV equation (\ref{e:ekdv}) take the water wave values.  Previous 
work to explain this minimal amplitude based on resonant solitary wave theory predicted that this amplitude vanishes for a fixed combination of the higher order coefficients, with minimal amplitude in a neighbourhood of this node \cite{resekdv}.  However, the present modulation theory predicts that the resonant wave amplitude is minimal for regions in the higher order coefficient parameter space, as also shown by full numerical solutions of the eKdV equation.

The combination of modulation theory with the concept of a Whitham shock and approximate theory can be used for other problems which involve resonant dispersive shock waves.  These include resonant optical dispersive shock waves in nematic liquid crystals \cite{saleh,nonlocaltolocal,salehthesis}.  These are governed by the eKdV equation for small steps in the optical beam power which generates them, but for general initial steps the governing equations are more complicated, consisting of a nonlinear Schr\"odinger-type equation for the optical field and an elliptic equation for the nematic response \cite{PR}.  

\section*{Acknowledgement}
S.B. would like to thank Patrick Sprenger with whom this research work stimulated extensive discussions. S.B. is grateful for the hospitality of Westminster College arranged by the Isaac Newton Institute (INI) for Mathematical Sciences, University of Cambridge, where part of this work was reported and discussed in the Physical Applications Workshop HY2W05. The authors are thankful for the referees' comments and suggestions which greatly improved the manuscript.

\appendix

\section{CDSW equal amplitude relation}
\label{a:appendixa}

The averaged mass and energy densities for the eKdV solitary wave are, respectively,
\begin{equation}
    \int_{-\infty}^{\infty} u_{s} \: dx = \bar{u}_{s}+2\sqrt{2}\sqrt{a_{s}}+\epsilon\sqrt{2} a^{3/2}_{s}\left(2c_{6}+\frac{4}{3}c_{7}\right),
    \label{e:masssol}
\end{equation}
\begin{eqnarray}
     & & \int_{-\infty}^{\infty} \left[ \frac{1}{2} u_{s}^{2} - \frac{1}{3} \epsilon \left( c_{2} - \frac{1}{2} c_{2} \right)u_{s}^{3} \right] \: dx= \frac{1}{2} \bar{u}^2_{s} + 2\sqrt{2}\bar{u}_{s}\sqrt{a_s} + \frac{2\sqrt{2}}{3}a^{3/2}_s \nonumber \\
     & & \mbox{} + \epsilon \left[ \frac{1}{4} \left(c_2 - 2c_3 \right) \bar{u}_{s}^{3}
     + \frac{3\sqrt{2}}{2} \sqrt{a_s}\left( c_2 - 2c_3 \right)\bar{u}_{s}^{2} + \sqrt{2} a^{3/2}_s\left( c_{2} - 2c_{3} + 2c_{6} + \frac{4}{3} c_7
     \right) \bar{u}_{s} \right. \nonumber 
     \\ 
     & & \left.\mbox{} + \frac{4\sqrt{2}}{15} \left( 5c_{6} + 4c_7\right) a_{s}^{5/2} \right].
     \label{e:energysol}
\end{eqnarray}

The averaged mass and energy fluxes for the resonant Stokes wave are, respectively,
\begin{equation}
    \bar{Q}_{mr} = 3\bar{u}^2_{r}+\frac{3}{2}a^2_r +\epsilon \left[ \frac{1}{3} c_1\bar{u}^{3}_{r} + \frac{1}{2} c_1\bar{u}_{r}a_{r}^{2} + \frac{1}{4}c_{2} k^2_r - \frac{3}{4}c_3k^2_r \right] ,
    \label{e:fluxmass} 
    \end{equation}
\begin{eqnarray}
\bar{Q}_{er} & = & 2\bar{u}^{3}_{r} + 3 \left(\bar{u}_{r} - \frac{1}{4} k^{2}_{r}\right)a_{r}^{2} + \epsilon \left[ \frac{1}{4} \left(c_1 + 3c_{2} - 6c_3 \right) \bar{u}_{r}^{4} \right.  \nonumber\\ 
& & \left. \mbox{} + \frac{1}{4} \left( \left( 3c_1 + 9c_{2} - 18c_{3}\right)\bar{u}^{2}_{r} - 2c_2\bar{u}_{r} k^2_r + 5c_{4}k_{r}^{4} \right) a_{r}^{2} \right].
\label{e:fluxenergy}
\end{eqnarray}

\section{Modulation theory jump conditions}
\label{a:jump}

\setcounter{equation}{0}
\renewcommand{\theequation}{B \arabic{equation}}

The averaged mass density for the resonant Stokes wave and the bore in the CDSW regime to $O(\epsilon)$ are, respectively, 
\begin{equation}
    \bar{P}_{mr} = \bar{u}_{r},
    \label{e:pmr}
\end{equation}
\begin{equation}
    \bar{P}_{mcdsw} = \bar{u}_{s}+2\sqrt{2}\sqrt{a_s} + 2\sqrt{2} \epsilon \left(c_{6}+\frac{2}{3}c_{7}\right) a_{s}^{3/2}.
    \label{e:pmcdsw}
\end{equation}
The averaged mass flux for the resonant Stokes wave and the bore in the CDSW regime to $O(\epsilon)$ are, respectively, 
\begin{equation}
    \bar{Q}_{mr} = 3\bar{u}^2_{r}+\frac{3}{2}a^2_r +\epsilon \left[ \frac{1}{3} c_1\bar{u}^{3}_{r} + \frac{1}{2} c_1\bar{u}_{r}a_{r}^{2} + \frac{1}{4}c_{2} k^2_r - \frac{3}{4}c_3k^2_r \right] 
    \label{e:qmr}
\end{equation}
and
\begin{eqnarray}
& & \bar{Q}_{mcdsw} = 3\bar{u}_{s}^2 + 12\sqrt{2}\bar{u}_{s}\sqrt{a_s} + 4\sqrt{2} a^{3/2}_s + \epsilon\left[ \frac{1}{3}\bar{u}_s^3 c_1 + 2\sqrt{2}c_{1} \bar{u}^2_s\sqrt{a}_{s} + 4\sqrt{2} \left( \frac{1}{3} c_{1} \right. \right. \nonumber \\
& & \left. \left. \mbox{} + 3c_6 + 2c_{7} \right)\bar{u}_{s}a_{s}^{3/2} + 4\sqrt{2}\left( \frac{1}{15} c_{2} -\frac{1}{5}c_{3} + 2c_{6} + \frac{8}{5} c_{7}  \right) a_{s}^{5/2} \right].
\label{e:qmcdsw}
\end{eqnarray}

The averaged energy density for the resonant Stokes wave and the bore in the CDSW regime, with all terms taken into account for the CDSW, as discussed in Section \ref{s:modulation}, are, respectively,
\begin{eqnarray}
\bar{P}_{er} & = & \frac{1}{2}\bar{u}^2_r +\frac{1}{4}a^2_{r}+\frac{1}{4}u^2_{2}a^{4}_{r} + \frac{1}{6}\epsilon \left(c_2 - 2c_3\right)\bar{u}_{r}^{3} + \frac{1}{4}\epsilon \left(c_2 - 2c_{3}\right)\bar{u}_{r}a_{r}^{2}  \nonumber \\ 
& &  \mbox{} + \frac{1}{8} \epsilon \left(c_{2} u_2 + 2c_{2}u_{2}^{2}\bar{u}_{r} - 2c_{3}u_{2} - 4c_{3} u_{2}^{2}\bar{u}_{r} \right)a_{r}^{4}
\label{e:per}
\end{eqnarray}
and
\begin{equation}
\bar{P}_{ecdsw} = \bar{P}_{ecdsw,0} + \epsilon \bar{P}_{ecdsw,1} + \epsilon^{2} \bar{P}_{ecdsw,2} + \epsilon^{3} \bar{P}_{ecdsw,3} + \epsilon^{4} \bar{P}_{ecdsw,4},
\label{e:pbar}
\end{equation}
where
\begin{eqnarray}
 & & \bar{P}_{ecdsw,0} = \frac{1}{2}\bar{u}^2_s + 2\sqrt{2}\bar{u}_{s}\sqrt{a}_{s}+\frac{2}{3}\sqrt{2}a^{3/2}_{s}, \nonumber \\
 & & \bar{P}_{ecdsw,1} = \frac{1}{6}\left(c_{2}-2c_{3}\right)\bar{u}_{s}^{3} + \sqrt{2}\left(c_2-2c_3\right)\bar{u}_{s}^{2}\sqrt{a_{s}} \nonumber \\ 
& &  \mbox{} + \frac{2}{3}\sqrt{2}\left(c_{2} - 2c_{3} + 3c_6 + 2c_7\right)\bar{u}_{s}a_{s}^{3/2} + \frac{4}{15}\sqrt{2}\left(5c_6 + 4c_7\right)a_{s}^{5/2}, \nonumber \\
& & \bar{P}_{ecdsw,2} = \sqrt{2}\left(c_2c_{6} + \frac{2}{3}c_2c_7-2c_3c_6-\frac{4}{3}c_3c_7\right)\bar{u}_{s}^{2}a_{s}^{3/2}  + \frac{4}{15}\sqrt{2}\left(5c_2c_6 
+ 4c_2c_7 \right. \nonumber \\
& & \left. \mbox{} - 10c_3c_6 - 8c_3c_7 \right)
\bar{u}_{s}a_{s}^{5/2} + 2\sqrt{2}\left( \frac{4}{15}c_{2}c_{6} + \frac{8}{35}c_{2}c_{7} - \frac{8}{15}c_{3}c_{6} - \frac{16}{35}c_{3}c_{7} + \frac{1}{3}c^2_6 \right. \nonumber \\
& & \left. \mbox{} + \frac{8}{15}c_6c_7
+ \frac{4}{17}c_{7}^{2} \right)a_{s}^{7/2}, \label{e:pecdsw} \\
& & \bar{P}_{ecdsw,3} = 2\sqrt{2}\left( \frac{1}{3}c_{2}c_{6}^{2} +  \frac{8}{15}c_2c_6c_7 + \frac{8}{35}c_{2}c_{7}^{2} - \frac{16}{15}c_3c_6c_7- \frac{2}{3}c_3c^2_6 \right. \nonumber \\
& & \left. \mbox{} - \frac{16}{35}c_{3}c_{7}^{2} \right)\bar{u}_{s}a_{s}^{7/2} + \frac{8}{5}\sqrt{2}\left( \frac{1}{3}c_2c^2_6 - \frac{2}{3}c_3c^2_6 + \frac{4}{7}c_2c_6c_7 - \frac{8}{7}c_3c_6c_7 \right)a_{s}^{9/2}, \nonumber \\
& & \bar{P}_{ecdsw,4} = 8\sqrt{2} \left[ \frac{1}{45}c_{2}c_{6}^{3} + \frac{2}{35}c_{2}c_{6}^{2}c_{7} + \frac{16}{315}c_{2}c_{6}c_{7}^{2} + \frac{32}{2079}c_{2}c_{7}^{3} - \frac{2}{45}c_{3}c_{6}^{3} - \frac{4}{35}c_{3}c_{6}^{2}c_{7} \right. \nonumber \\ 
& & \left. \mbox{} -
\frac{32}{315}c_3c_6c^2_7 - \frac{64}{2079}c_3c^3_7 \right] a_{s}^{11/2}. \nonumber 
\end{eqnarray}

The averaged energy flux for the resonant Stokes wave and the bore in the CDSW regime, with all terms taken into account for the CDSW, as discussed in Section \ref{s:modulation}, are, respectively,
\begin{eqnarray}
\bar{Q}_{er} & = & 2\bar{u}_{r}^{3} + 3\left(\bar{u}_r -\frac{1}{4}k^2_r \right)a_{r}^{2} + 3\left(\frac{1}{2} - k^2_ru_2 + u_{2}\bar{u}_r \right) u_{2}a_{r}^{4} + \frac{1}{4}\epsilon \left( 3c_{1}\bar{u}^2_r + 9c_{2}\bar{u}^2_r \right.  \nonumber\\ 
& &  \mbox{} \left. - 2c_{2}\bar{u}_rk^2_r - 18c_{3}\bar{u}^2_r + 5c_{4}k^4_r \right)a_{r}^{2} + \frac{1}{32}\epsilon \left( 3c_{1} + 24c_{1}u_{2}\bar{u}_{r} + 24c_{1}u_{2}^{2}\bar{u}_{r}^{2} + 9c_{2} \right.  \nonumber\\ 
& &  \left. \mbox{} + 72c_{2}u_{2}^{2}\bar{u}^2_r + 72c_{2}u_{2}\bar{u}_r - 24c_{2}u_{2}k^2_r - 64c_{2}k^2_ru^2_2\bar{u}_r - 18c_{3} - 144c_{3}u_{2}^{2}\bar{u}^2_r - 144c_{3}u_{2}\bar{u}_r  \right.  \nonumber\\ 
& &  \left. \mbox{} + 640c_{4}k^4_ru^2_2  \right)a_{r}^{4} + \frac{3}{8}\epsilon \left( c_{1} + 3c_2 - 6c_3 \right)u_{2}^{2}a_{r}^{6} 
+ \frac{3}{32}\epsilon \left( c_1 + 3c_2 - 6c_3 \right)u_{2}^{4}a_{r}^{8} ,
\label{e:qer}
\end{eqnarray}
and
\begin{equation}
 \bar{Q}_{ecdsw} = \bar{Q}_{ecdsw,0} + \epsilon \bar{Q}_{ecdsw,1} + \epsilon^{2} \bar{Q}_{ecdsw,2}
 + \epsilon^{3} \bar{Q}_{ecdsw,3} + \epsilon^{4} \bar{Q}_{ecdsw,4} + \epsilon^{5}\bar{Q}_{ecdsw,5},
\end{equation}
where
\begin{eqnarray}
 & & \bar{Q}_{ecdsw,0} = 2\bar{u}^{3}_{s} + 12\sqrt{2}\bar{u}^{2}_{s}\sqrt{a_s} + 8\sqrt{2}\bar{u}_{s}a^{3/2}_s - \frac{4}{5}\sqrt{2}a^{5/2}_{s}, \nonumber \\
& &  \bar{Q}_{ecdsw,1} = \frac{1}{4} \left( c_{1} + 3c_2 - 6c_3\right) \bar{u}^{4}_{s} + 2\sqrt{2} \left( c_1+3c_2-6c_3\right) \bar{u}^{3}_{s}\sqrt{a_s} + 2\sqrt{2} \left(c_{1} + 3c_{2} - 6c_{3} \right. \nonumber \\
 & & \left. \mbox{} + 6c_{6} + 4c_7 \right) \bar{u}^2_s a^{3/2}_s + 8\sqrt{2}\left(-\frac{1}{15} c_{2} + 2c_{6} + \frac{8}{5}c_7 \right) \bar{u}_s a^{5/2}_s 
 + 8\sqrt{2} \left(\frac{1}{21}c_4  + \frac{3}{5}c_6 + \frac{16}{35}c_7\right) a^{7/2}_s, \nonumber \\
 & & \bar{Q}_{ecdsw,2} = 2\sqrt{2} \left(\frac{2}{3}c_1c_7 + c_1c_6 + 2c_2c_7 + 3c_2c_6 - 4c_3c_7 - 6c_3c_6   
\right) \bar{u}^3_s a^{3/2}_{s} 
\nonumber \\
& & \mbox{} + 4\sqrt{2}\left( c_{1}c_{6} + \frac{4}{5}c_1c_7 + 3c_2c_6 + \frac{12}{5}c_2c_7 - 6c_{3}c_{6} - \frac{24}{5}c_3c_7
\right) \bar{u}^2_s a^{5/2}_{s} 
\nonumber \\
& & \mbox{} + \frac{8}{5}\sqrt{2}\left( 2c_{1}c_{6} + \frac{12}{7}c_{1}c_{7} + \frac{16}{3}c_{2}c_{6} + \frac{92}{21}c_{2}c_{7} - \frac{72}{7}c_{3}c_{7} - 12c_{3}c_{6} + 5c_{6}^{2} + 8c_6c_7 \right. \nonumber \\
& & \left. \mbox{} + \frac{24}{7}c^2_7 \right) \bar{u}_{s}a^{7/2}_{s}
+ \frac{2}{5} \sqrt{2} \left( 5c_{1}c_{6}^{2} + \frac{24}{7}c_{1}c_{7}^{2} + 24c_{2}c_{6}c_{7} +  \frac{72}{7}c_2c^2_7 + 15c_2c^2_6 
- 48c_{3}c_{6}c_{7} \right. \nonumber \\
& & \left. \mbox{} - \frac{144}{7}c_3c^2_7 - 30c_3c^2_6 + 8c_1c_6c_7 
\right) \bar{u}^{2}_{s} a^{7/2}_{s} 
+ 4\sqrt{2} \left( -\frac{8}{35}c_{2}c_{6} - \frac{16}{63}c_{2}c_{7} + \frac{20}{21}c_{4}c_{6} + \frac{32}{21}c_{4}c_{7} \right. \nonumber \\
& & \left. \mbox{} + \frac{7}{5}c_{6}^{2} + \frac{16}{7}c_6c_7 
-\frac{32}{105}c^2_7 \right) a^{9/2}_{s}\bar{u}^{2}_{s}, \nonumber \\
& & \bar{Q}_{ecdsw,3} = 8\sqrt{2}\left(  \frac{2}{5}c_{1}c_{6}^{2} +\frac{32}{105}c_{1}c_{7}^{2} + \frac{24}{35}c_1c_6c_7 + \frac{17}{15}c_{2}c_{6}^{2} +\frac{40}{21}c_2c_6c_7 + \frac{256}{315}c_{2}c_{7}^{2}
\right. \nonumber \\
& & \left. \mbox{} -\frac{144}{35}c_3c_6c_7
- \frac{12}{5}c_3c^2_6 -\frac{64}{35}c_3c^2_7 \right) 
\bar{u}_{s}a^{9/2}_{s}
+ 8\sqrt{2} \left( \frac{6}{35}c_1c^2_6 + \frac{32}{231}c_1c^2_7 +\frac{32}{105}c_1c_6c_7 + \frac{2}{5}c_{2}c_{6}^{2} 
\right. \nonumber \\ 
& & \left. \mbox{} + \frac{208}{315}c_2c_6c_7 + \frac{928}{3465}c_{2}c_{7}^{2} - \frac{36}{35}c_{3}c_{6}^{2} - \frac{64}{77}c_{3}c_{7}^{2}   \right. 
\left. -\frac{64}{35}c_3c_6c_7 +\frac{5}{21}c_4c^2_6 
+ \frac{512}{693}c_{4}c_{7}^{2} \right. \nonumber \\ 
& & \left. \mbox{} + \frac{16}{21}c_4c_6c_7  \right. 
\left. + \frac{4}{15}c_{6}^{3} +\frac{24}{35}c^2_6c_7 + \frac{64}{105}c_6c^2_7 + \frac{128}{693}c^3_7 \right) a_{s}^{11/2}, \nonumber \\
& & \bar{Q}_{ecdsw,4} = 16\sqrt{2}\left( \frac{1}{15}c_{1}c_{6}^{3} + \frac{16}{35}c_{2}c_{6}c^{2}_{7} + \frac{16}{105}c_{1}c_{6}c_{7}^{2} + \frac{6}{35}c_{1}c_{6}^{2}c_{7}
+ \frac{32}{693} c_{1}c_{7}^{3} + \frac{1}{5}c_{2}c_{6}^{3} \right. \nonumber \\
& & \left. \mbox{} + \frac{18}{35}c_{2}c_{6}^{2}c_{7} 
+ \frac{32}{231}c_2c^3_7 -\frac{2}
{5}c_{3}c_{6}^{3} - \frac{64}{231}c_3c^3_7 
-\frac{32}{35}c_3c_6c^2_7 -\frac{36}{35}c_3c_{6}^{2}c_7
\right)\bar{u}_{s}a^{11/2}_{s} \nonumber \\
& & \mbox{} + 32\sqrt{2}\left(  \frac{1}{35}c_{1}c_{6}^{3} + \frac{8}{105}c_{1}c_{6}^{2}c_{7} + \frac{16}{315}c_{1}c_{6}^{3}c_{7} + \frac{64}{3003}c_{1}c_{7}^{3} + \frac{8}{105}c_{2}c_{6}^{3}
+ \frac{64}{1287}c_{2}c_{7}^{3} \right. 
\nonumber \\
& & \left. \mbox{} + \frac{62}{315}c_{2}c_{6}^{2}c_{7}  + \frac{16}{105}c_{2}c_{6}^{3}c_{7} - \frac{128}{3465}c_{2}c_{6}c_{7}^{2} - \frac{6}{35}c_3c^3_6 - \frac{16}{35}c_{3}c_{6}^{2}c_{7} - \frac{32}{105}c_3c^3_6c_7 \right. \nonumber \\
& & \left. \mbox{} - \frac{128}{1001}c_{3}c_{7}^{3} \right) a_{s}^{13/2}, \nonumber \\
& & \bar{Q}_{ecdsw,5} = 8\sqrt{2}\left( \frac{1}{35}c_{1}c_{6}^{4} + \frac{32}{315}c_{1}c_{6}^{3}c_{7}
+ \frac{32}{231}c_{1}c_{6}^{2}c_{7}^{2} + \frac{256}{3003}c_{1}c_{6}c_{7}^{3} + \frac{128}{6435}c_{1}c_{7}^{4} + \frac{3}{35}c_2c^4_6 \right. \nonumber \\
& & \left. \mbox{} + \frac{32}{105}c_{2}c_{6}^{3}c_{7}
+ \frac{32}{77}c_{2}c_{6}^{2}c_{7}^{2} + \frac{256}{1001}c_{2}c_{6}c_{7}^{3}
+ \frac{128}{2145}c_{2}c_{7}^{4} - \frac{6}{35}c_{3}c_{6}^{4} - \frac{64}{105}c_{3}c_{6}^{3}c_{7} \right. \nonumber \\
& & \left. \mbox{} - \frac{64}{77}c_{3}c_{6}^{2}c_{7}^{2} - \frac{512}{1001}c_3c_6c^3_7 - \frac{256}{2145}c_{3}c_{7}^{4} \right) a_{s}^{15/2}.
\label{e:enfluxaverage}
\end{eqnarray}

\bibliographystyle{vancouver}
%\bibliography{research}

\begin{thebibliography}{99}

\bibitem{whitham}  G.B. Whitham, 1974, {\em Linear and Nonlinear Waves,} J. Wiley and Sons, New York.

%\bibitem{benjamin}  T.B. Benjamin and M.J. Lighthill, 1954, ``On cnoidal waves and bores,'' 
%{\em Proc.\ Roy.\ Soc.\ Lond.\ A,} {\bf 224}, 448--460.
%
%\bibitem{johnson70} R.S. Johnson, 1970, ``A non-linear equation incorporating damping and dispersion,'' 
%{\em J. Fluid Mech.,} {\bf 42}, 49--60.

\bibitem{baines}  P.G. Baines, 1995, {\em Topographic Effects in Stratified Flows,}
Cambridge Monographs on Mechanics, Cambridge. 
     
\bibitem{esler}  J.G. Esler and J.D. Pearce, 2011, ``Dispersive dam-break and lock-exchange flows in a two-layer fluid,'' 
{\em J. Fluid Mech.,} {\bf 667}, 555--585.

\bibitem{christie}  D.R. Christie, 1989, ``Long nonlinear waves in the lower atmosphere,''
{\em J. Atmos.\ Sci.,} {\bf 46}, 1462--1491.

\bibitem{clarke}  R.H. Clarke, R.K. Smith and D.G. Reid, 1981, ``The morning glory of
the Gulf of Carpentaria: an atmospheric undular bore,'' {\em Monthly
Weather Rev.,} {\bf 109}, 1726--1750.

\bibitem{anne}  V.A. Porter and N.F. Smyth, 2002, ``Modelling the Morning Glory of
     the Gulf of Carpentaria,'' {\em J. Fluid Mech.,} {\bf 454}, 1--20.
     
\bibitem{nwshelf}  N.F. Smyth and P.E. Holloway, 1988, ``Hydraulic jump and undular bore
formation on a shelf break,'' {\em J. Phys.\ Ocean.,} {\bf 18}, 947--962.
     
%\bibitem{resflow}  R.H.J. Grimshaw and N.F. Smyth, 1986, ``Resonant flow of a stratified fluid over topography,''
%{\em J. Fluid Mech.,} {\bf 169}, 429--464.
% 
%\bibitem{procroy}  N.F. Smyth, 1987, ``Modulation theory solution for resonant flow over topography,'' {\em Proc.\
%Roy.\ Soc.\ Lond.\ A,} {\bf 409}, 79--97.
%
%\bibitem{esler}  J.G. Esler and J.D. Pearce, 2011, ``Dispersive dam-break and lock-exchange flows in a two-layer 
%fluid,'' {\em J. Fluid Mech.,} {\bf 667}, 555--585.

\bibitem{scott1}  D.R. Scott and D.J. Stevenson, 1984, ``Magma solitons,'' {\em Geophys.\ Res.\ Lett.,} 
{\bf 11}, 1161--1164.

\bibitem{scott2}  D.R. Scott and D.J. Stevenson, 1986, ``Magma ascent by porous flow,'' 
{\em Geophys.\ Res.\ Lett.,} {\bf 91}, 9283--9296.

\bibitem{hoefer2}  N.K. Lowman and M.A. Hoefer, 2013, ``Dispersive shock waves in viscously deformable media,'' 
{\em J. Fluid Mech.,} {\bf 718}, 524--557.

%\bibitem{magma}  T.R. Marchant and N.F. Smyth, 2005, ``Approximate solutions for
%     magmon propagation from a reservoir,'' {\em IMA J. Appl.\ Math.,} {\bf 70}, 796--813.

\bibitem{hooper} C.G. Hooper, P.D. Ruiz, J.M. Huntley and K.R. Khusnutdinova, 2021, ``Undular bores generated by fracture,'' {\em Phys.\ Rev.\ E,} {\bf 104}, 044207.
     
\bibitem{fleischer2}  C. Barsi, W. Wan, C. Sun and J.W. Fleischer, 2007, ``Dispersive shock waves with nonlocal nonlinearity,'' {\em Opt.\ Lett.,} {\bf 32}, 2930--2932.

\bibitem{fleischer}  W. Wan, S. Jia and J.W. Fleischer, 2007, ``Dispersive superfluid-like shock waves in nonlinear optics,'' {\em Nature Phys.,} {\bf 3}, 46--51.

\bibitem{elopt}  G.A. El, A. Gammal, E.G. Khamis, R.A. Kraenkel and A.M. Kamchatnov, 2007,
``Theory of optical dispersive shock waves in photorefractive media,'' {\em Phys.\ Rev.\ A,} {\bf 76}, 053183.

\bibitem{colloid}  X. An, T.R. Marchant and N.F. Smyth, 2017, ``Optical dispersive shock waves in defocusing colloidal media,'' {\em Physica D,} {\bf 342}, 45--56.

\bibitem{nemboreel}  G. El and N.F. Smyth, 2016, ``Radiating dispersive shock waves in non-local optical media,'' {\em Proc.\ Roy.\ Soc.\ Lond.\ A,} {\bf 472}, 20150633.

\bibitem{nemborephysd}  N.F. Smyth, 2016, ``Dispersive shock waves in nematic liquid crystals,'' {\em Physica D}, {\bf 333}, 301--309.

\bibitem{saleh}  S. Baqer and N.F. Smyth, 2020, ``Modulation theory and resonant regimes for dispersive shock waves in nematic liquid crystals,'' {\em Physica D,} {\bf 403}, 132334.

\bibitem{nonlocaltolocal}  S. Baqer, D.J. Frantzeskakis, T.P. Horikis, C. Houdeville, T.R. Marchant and N.F. Smyth, 2021, ``Nematic dispersive shock waves from nonlocal to local,'' {\em Appl.\ Sci.,} {\bf 11,} 4736.

\bibitem{salehthesis} S. Baqer, 2020, ``Dispersive hydrodynamics in a non-local non-linear medium,'' Ph.D. thesis, University of Edinburgh.

\bibitem{bose}  G.A. El, A.M. Kamchatnov, V.V. Khodorovskii, E.S. Annibale and A. Gammal, 2009,
``Two-dimensional supersonic nonlinear Schr\"odinger flow past an extended obstacle,'' {\em Phys.\ Rev.\ E,} {\bf 80}, 046317.

\bibitem{hoefer1}  N.K. Lowman and M.A. Hoefer, 2013, ``Fermionic shock waves:  Distinguishing dissipative
versus dispersive resolutions,'' {\em Phys.\ Rev.\ A,} {\bf 88}, 013605.

\bibitem{mod1}  G.B. Whitham, 1965, ``A general approach to linear and non-linear dispersive waves using a 
Lagrangian,'' {\em J. Fluid Mech.,} {\bf 22}, 273--283.

\bibitem{modproc}  G.B. Whitham, 1965, ``Non-linear dispersive waves,'' {\em Proc.\ Roy.\ Soc.\ London 
A,} {\bf 283}, 238--261.

\bibitem{BF} T. B. Benjamin and J. E. Feir, 1967, ``The disintegration of wave trains in deep water. Part 1. Theory,'' {\em J. Fluid Mech.,} {\bf 27},  417--430.

\bibitem{gur}  A.V. Gurevich and L.P. Pitaevskii, 1974, ``Nonstationary structure of a collisionless shock wave,'' 
{\em Sov.\ Phys.\ JETP,} {\bf 33}, 291--297.

\bibitem{bengt}  B. Fornberg and G.B. Whitham, 1978, ``Numerical and theoretical study of certain non-linear wave 
phenomena,'' {\em Phil.\ Trans.\ Roy.\ Soc.\ Lond.\ Ser.\ A,} {\bf 289}, 373--404.

\bibitem{flash}  H. Flaschka, M.G. Forest and D.W. McLaughlin, 1980, ``Multiphase averaging and the inverse 
spectral solution of the Korteweg-de Vries equation,'' {\em Comm.\ Pure Appl.\ Math.,} {\bf 33}, 739--784.

\bibitem{el2}  G.A. El, 2005, ``Resolution of a shock in hyperbolic systems modified by 
weak dispersion,'' {\em Chaos,} {\bf 15}, 037103.

\bibitem{elreview} G.A. El and M.A. Hoefer, 2016, ``Dispersive shock waves and modulation theory,'' 
{\em Physica D,} \textbf{333}, 11--65.

\bibitem{kawahara} T. Kawahara, 1972, ``Oscillatory solitary waves in dispersive
media,'' {\em J. Phys.\ Soc.\ Japan,} {\bf 33}, 260--264.

\bibitem{boyd}  J.P. Boyd, 1991, ``Weakly non-local solutions for capillary-gravity waves: fifth degree Korteweg-de 
Vries equation,'' {\em Physica D,} {\bf 48}, 129--146.

\bibitem{grim93} E.S. Benilov, R. Grimshaw and E.P. Kuznetsova, 1993, 
``The generation of radiating waves in a singularly-perturbed Korteweg-de Vries equation,'' {\em Physica D,}  
{\bf 69}, 270--278.

\bibitem{radiating_sol1} R. Grimshaw, B. Malomed and E.S. Benilov, 1994, ``Solitary waves with 
damped oscillatory tails: an analysis of the fifth-order Korteweg-de Vries equation,'' {\em Physica D,} 
{\bf 77}, 473--485.

\bibitem{radiating_sol2} V.I. Karpman, 1998, ``Radiation by weakly nonlinear shallow-water solitons due to 
higher-order dispersion,'' {\em Phys.\ Rev.\ E,} {\bf 58}, 5070--5080. 

\bibitem{kivshar} V.V. Afanasjev, Y.S. Kivshar and C.R. Menyuk, 1996, ``Effect of third-order dispersion on 
dark solitons,'' {\em Opt.\ Lett.,} {\bf 21}, 1975--1977.

\bibitem{markkaw}  P. Sprenger and M.A. Hoefer, 2017, ``Shock waves in dispersive hydrodynamics with non-convex dispersion,'' {\em SIAM J. Appl.\ Math.,} {\bf 77}, 26--50.

\bibitem{resekdv}  T.P. Horikis, D.J. Frantzeskakis, T.R. Marchant and N.F. Smyth, 2021, ``Higher-dimensional extended shallow water equations
and resonant soliton radiation,'' {\em Phys.\ Rev.\ Fluids,} {\bf 6}, 104401.

\bibitem{trillores}  M. Conforti and S. Trillo, 2013, ``Dispersive wave emission from wave breaking,'' {\em Opt.\ Lett.,} {\bf 38}, 3815--3818.

\bibitem{trillo1}  M. Conforti, F. Baronio and S. Trillo, 2014, ``Resonant radiation shed by dispersive
shock waves,'' {\em Phys.\ Rev.\ A,} {\bf 89}, 013807.

\bibitem{trilloresfour}  M. Conforti and S. Trillo, 2014, ``Radiative effects driven by shock waves in cavity-less four-wave mixing combs,'' {\em Opt.\ Lett.,} {\bf 39}, 5760--5763.

\bibitem{trilloresnature}  M. Conforti, S. Trillo, A. Mussot and A. Kudlinski, 2015, ``Parametric excitation of multiple resonant radiations from localized wavepackets,'' {\em Sci.\ Rep.,} {\bf 5}, 1--5.

\bibitem{trilloreslossbore}  S. Malaguti, M. Conforti and S. Trillo, 2014, ``Dispersive radiation induced by shock waves in passive resonators,'' {\em Opt.\ Lett.,} {\bf 39}, 5626--5629.

\bibitem{ekdv}  T.R. Marchant and N.F. Smyth, 1990, ``The extended Korteweg-de Vries equation and the resonant flow of a fluid over topography'', {\em J. Fluid Mech.,} {\bf 221}, 263--288.

\bibitem{tovbis}  G.A. El, E.G. Khamis and A. Tovbis, 2016, ``Dam break problem for the focusing nonlinear Schr\"odinger 
equation and the generation of rogue waves,'' {\em Nonlinearity,} {\bf 29}, 2798--2836.

\bibitem{gennadygas} G.A. El, 2021, ``Soliton gas in integrable dispersive hydrodynamics,'' {\em J. Stat.\ Mech.,} {\bf 2021}, 114001.

\bibitem{thibaultgas} T. Congy, G.A. El, and G. Roberti, 2021, ``Soliton gas in bidirectional dispersive hydrodynamics,'' {\em Phys.\ Rev.\ E,} {\bf 103}, 042201.

\bibitem{vandyke}  M. Van Dyke, 1982, {\em An Album of Fluid Motion,} Parabolic Press, Stanford, California.

\bibitem{perturbkdv}  T.R. Marchant and N.F. Smyth,
2006, ``An undular bore solution for the higher-order Korteweg-de Vries equation,'' {\em J. Phys.\ A: Math.\ Gen.,} {\bf 39}, L563--569.

\bibitem{pat}   M.A. Hoefer, N.F. Smyth and P. Sprenger, 2019, ``Modulation theory solution for nonlinearly
resonant, fifth-order Korteweg-de Vries, nonclassical, traveling dispersive shock waves,'' {\em Stud.\ 
Appl.\ Math.,} {\bf 142}, 219--240.

\bibitem{patjump}  P. Sprenger and M.A. Hoefer, 2020, ``Discontinuous shock solutions of the Whitham modulation equations and traveling wave solutions of higher order dispersive nonlinear wave equations,'' {\em Nonlinearity,} {\bf 33}, 3268--3302.


\bibitem{sergeyshockfront}  S. Gavrilyuk, B. Nkonga, K. Shyue and L. Truskinovsky, 2020, ``Stationary shock-like transition fronts in dispersive systems,'' {\em Nonlinearity,} {\bf 33},  5477--5509.

\bibitem{sergeybbm}  S. Gavrilyuk and K. Shyue, 2022, ``Singular solutions of the BBM equation: analytical and numerical study,'' {\em Nonlinearity,} {\bf 35},  388--410.

\bibitem{boreapprox}  T.R. Marchant and N.F. Smyth, 2012, ``Approximate techniques for dispersive shock 
waves in nonlinear media,'' {\em J. Nonlin.\ Opt.\ Phys.\ Mater.,} {\bf 21}, 1250035.

\bibitem{negative}  T.R. Marchant and N.F. Smyth, 2002, ``The initial-boundary problem for the Korteweg-de Vries equation on the negative quarter-plane,'' {\em Proc.\ Roy.\ Soc.\ Lond.\ A,} {\bf 458}, 857--871 .

\bibitem{chan}  T. F. Chan and T. Kerkhoven, 1985, ``Fourier methods with extended stability
intervals for KdV,'' {\em SIAM J. Numer.\ Anal.,}  {\bf 22}, 441--454.

\bibitem{tref}  L. N. Trefethen, 2000, {\em Spectral Methods in MATLAB}, SIAM, Philadephia.

\bibitem{PR}  M. Peccianti and G. Assanto, 2012, ``Nematicons,'' {\em Phys.\ Rep.,} {\bf 516}, 147--208.

%\bibitem{fleischer3}  W. Wan, D.V. Dylov, C. Barsi and J.W. Fleischer, 2010, ``Diffraction from an edge on a
%self-focusing medium,'' {\em Opt.\ Lett.,} {\bf 35}, 2819--2821.    

%\bibitem{trillo5}  C. Conti, A. Fratalochhi, M. Peccianti, G. Ruocco and S. Trillo, 2009, ``Observation of a
%gradient catastrophe generating solitons,'' {\em Phys.\ Rev.\ Lett.,} {\bf 102}, 083902.

%\bibitem{trillo4}  M. Crosta, A. Fratalocchi and S. Trillo, 2012, ``Double shock dynamics induced by the saturation
%of defocusing nonlinearities,'' {\em Opt.\ Lett.,} {\bf 37}, 1634--1636.

%\bibitem{trillo7}  J. Fatome, C. Finot, G. Millot, A. Armaroli and S. Trillo, 2014,
%``Observation of optical undular bores in multiple four-wave mixing,'' {\em Phys.\ Rev.\ X,} {\bf 4}, 021022.

%\bibitem{trillo6}  N. Ghofraniha, C. Conti, G. Ruocco and S. Trillo, 2007, ``Shocks in nonlocal media,'' {\em Phys.\
%Rev.\ Lett.,} {\bf 99}, 043903.

%\bibitem{focbore2}  T.R. Marchant and N.F. Smyth, 2012, ``Semi-analytical solutions for 
%dispersive shock waves in colloidal media,'' {\em J. Phys.\ B: Atomic, Molec.\ 
%Opt.\ Phys.,} {\bf 45}, 145401.    

%\bibitem{mod2}  G.B. Whitham, 1967, ``Variational methods and applications to water waves,'' 
%{\em Proc.\ Roy.\ Soc.\ Lond.\ A,} {\bf 299}, 6--25.

%\bibitem{cole}  J. Kevorkian and J.D. Cole, 1981, {\em Perturbation Methods in Applied Mathematics,} Springer-Verlag,
%New York.

%\bibitem{el1} G.A. El, V.V. Khodorovskii and A.V. Tyurina, 2003, ``Determination of boundaries of unsteady oscillatory zone 
%in asymptotic solutions of non-integrable dispersive wave equations,'' {\em Phys.\ Lett.\ A,} {\bf 318},
%526--536.

%\bibitem{gennady}  G.A. El, V.V. Geogjaev, A.V. Gurevich and A.L. Krylov, 1995, ``Decay of 
%an initial discontinuity in the defocusing NLS hydrodynamics,'' {\em Physica D,} 
%{\bf 87}, 186--192.

%\bibitem{ressolhnls}  K.E. Webb, Y.Q. Xu, M. Erkintalo and S.G. Murdoch, 2013, ``Generalized dispersive wave
%emission in nonlinear fiber optics, {\em Opt.\ Lett.,} {\bf 38}, 151--153.

%\bibitem{if}  F. If, P. Berg, P. L. Christiansen and O. Skovgaard, 1987, ``Split-step spectral method for nonlinear
%Schr\"odinger equation with absorbing boundaries,'' {\em J. Comp.\ Phys.,} {\bf 72}, 501--503.

%\bibitem{numrec}  W.H. Press, S.A. Teukolsky, W.T. Vetterling,
%and B.P. Flannery, 1992, {\em Numerical Recipes in Fortran.  The Art of
%Scientific Computing,} Cambridge University Press.

\end{thebibliography}

\end{document}